\documentclass{mn2e}
\usepackage{epsfig}
\usepackage{graphicx}
\newif\ifAMStwofonts
\AMStwofontstrue

\title{Far-IR Detection Limits $-$ II. Probing Confusion Including Source Confusion}

\author[W.-S.~Jeong et al.]
    {Woong-Seob~Jeong,$^{1,2}$\thanks{Woong-Seob~Jeong\,(jeongws@ir.isas.jaxa.jp)}
     Chris P.~Pearson,$^2$$^{,4}$\thanks{Further information contact Chris Pearson\,(cpp@ir.isas.jaxa.jp) \it{http://www.ir.isas.jaxa.jp/$\sim$cpp/astrof/}}
     Hyung Mok~Lee,$^1$ Soojong~Pak,$^3$ \newauthor and Takao~Nakagawa$^2$ \\
    $^1$ Department of Physics and Astronomy, Seoul National University, Shillim-Dong, Kwanak-Gu, Seoul 151-742, South Korea \\
    $^2$ Institute of Space and Astronautical Science, Japan Aerospace Exploration
    Agency, \\ ~~Yoshinodai 3-1-1, Sagamihara, Kanagawa 229-8510, Japan \\
    $^3$ Department of Astronomy and Space Science, Kyung Hee University, Yongin-Si, Gyeonggi-Do
    446-701, South Korea\\
    $^4$ European Space Astronomy Centre (ESAC), Apartado 50727, 28080 Madrid, Spain}

\date{Accepted .\\
      Received ;\\
      in original form 2003 August}

\pagerange{\pageref{firstpage}--\pageref{lastpage}}

\begin{document}

\label{firstpage}

\maketitle

\begin{abstract}

We present a comprehensive analysis for the determination of the confusion
levels for the current and the next generation of far-infrared surveys assuming
three different cosmological evolutionary scenarios. We include an extensive
model for diffuse emission from infrared cirrus in order to derive absolute
sensitivity levels taking into account the source confusion noise due to point
sources, the sky confusion noise due to the diffuse emission, and instrumental
noise. We use our derived sensitivities to suggest best survey strategies for
the current and the future far-infrared space missions \textit{Spitzer},
\textit{AKARI} (\textit{ASTRO-F}), \textit{Herschel}, and \textit{SPICA}. We
discuss whether the theoretical estimates are realistic and the competing
necessities of reliability and completeness. We find the best estimator for the
representation of the source confusion and produce predictions for the source
confusion using far-infrared source count models. From these confusion limits
considering both source and sky confusions, we obtain the optimal, confusion
limited redshift distribution for each mission. Finally, we predict the Cosmic
Far-Infrared Background (CFIRB) which includes information about the number and
distribution of the contributing sources.

\end{abstract}

\bigskip


\bigskip

\section{INTRODUCTION}\label{sec:introduction}

The detection limits of space infrared (IR) telescope systems are governed by
instrumental, photon, and confusion noise sources. The instrumental noise is
intrinsic to the detector and electronics system, e.g., readout noise and dark
current fluctuations. The photon noise is due to Poissonian fluctuations in the
photon counts from the sky background emission and the telescope thermal
emission which can be reduced by decreasing the telescope temperature. On the
other hand, the confusion noise is due to both the superposition of point
sources in crowded fields and the brightness fluctuation of extended structures
at scales of order of the telescope beam size. In the far-IR bands, due to the
relatively large beam sizes, astronomical observations are mostly affected by
the sky confusion due to the Galactic cirrus structure \cite{gautier92} and the
source confusion due to the unresolved extragalactic sources.

Though the confusion level is the unavoidable limit for the detection of faint
point sources, the source structure below the confusion limit also creates
background fluctuation in the observed image. This means that the source
distribution can be obtained by measuring the background fluctuation which
results from a convolution of the background sources and the telescope optics
\cite{lagache00b,matsuhara00}. After the successful  Infrared Astronomy
Satellite mission (\textit{IRAS}) \cite{soifer87}, subsequent space IR
telescopes such as the Infrared Space Observatory (\textit{ISO})
\cite{kessler96} and \textit{Spitzer} \cite{gall02,werner04}, have increased
the levels of sensitivity and spatial resolution at far-infrared wavelengths.
In addition, the Japanese \textit{AKARI} (formerly known as \textit{ASTRO-F})
satellite was launched successfully on Februrary 21st 2006
\cite{mura98,naka01,shib04,cpp04,cpp06b}, and will perform an all sky survey in
4 far-IR bands with much improved sensitivities, spatial resolutions and wider
wavelength coverage than the previous \textit{IRAS} survey over two decades
ago. Moreover, within the next decade, the Herschel Space Observatory
(\textit{Herschel}) \cite{pilb02,harwit04} and the Space Infrared Telescope for
Cosmology and Astrophysics (\textit{SPICA}) \cite{naka04} will observe the much
deeper universe with large aperture sizes of 3.5 m. In our previous paper
(Jeong et al. 2005, hereafter Paper I), we numerically generated a high spatial
resolution map of the Galactic cirrus and investigated the effect of sky
confusion for current and future space far-IR missions. In addition to the
simulated sky confusion noise, this work includes realistic source confusion
effects assuming various cosmological evolutionary scenarios. We also obtain
the expected Cosmic Far-IR Background (CFIRB).

The fluctuations in the surface brightness of extended structure on scales
smaller than the resolution of the telescope/instrument beam can produce
spurious events that can be easily mistaken for genuine point sources, since
the existence of a source is usually simply derived from the difference in
signal between the on source and some background (off source) position. Such
extended structure is observed in wispy neutral interstellar dust in the Milky
Way that is heated by the interstellar radiation field and is known as the
infrared cirrus~\cite{low84}, and is one of the main noise sources in the
far-IR range. Cirrus emission peaks at far-IR wavelengths but was detected in
all 4 \textit{IRAS} bands. There have been realistic estimations of the sky
confusion from observational data by \textit{IRAS} and \textit{ISO}
\cite{gautier92,HB90,herb98,kiss01,kiss05,jeong05}. The intensity of Galactic
cirrus is a function of Galactic latitude and is serious for wavelengths longer
than 60 $\mu$m. Using high resolution maps extended from currently available
Galactic emission maps, we have estimated the sky confusion noise for various
missions, based on fluctuation analysis and detailed photometry over
realistically simulated images. In Paper I, we concluded that the sky confusion
is expected to be two orders of magnitude lower for the next generation of
space missions with larger aperture sizes such as \textit{Herschel} and
\textit{SPICA}, but on the other hand, current 60--90 cm aperture missions such
as \textit{Spitzer} and \textit{AKARI}, will have to endure a sky dominated by
confusion noise at long wavelengths ($> 100~\mu$m). In this paper, we also
discuss the composite effect of sky confusion and source confusion from the
implementation of a realistic model for the source distribution.

Note that additional components affecting the confusion noise are the presence
of faint, unresolved, asteroids and the zodiacal light. Though bright asteroids
can be easily rejected via source confirmation between 2 pointings/scans over
the same sky position, faint asteroids may contribute to fluctuations assumed
to be from faint galaxies. Although $\acute{\rm A}$brah$\acute{\rm a}$m et al.
\shortcite{abra97} found no such fluctuations on small scales from the ISOPHOT
(photometer on board the \textit{ISO}) data, more sensitive instruments may
detect them. Thus, these two components may show a non-negligible contribution
to the far-IR confusion near the ecliptic plane.

This paper is structured as follows. In Section \ref{sec:confused}, we briefly
describe the confusion noise due to sky brightness fluctuations and the source
confusion due to extragalactic point sources. We present our source count
models including galaxy evolution and the simulated images in Section
\ref{sec:models}. Based upon the specifications of each IR mission, we estimate
the source confusion noise through fluctuation and photometric analyses in
Section \ref{sec:simulation}. We present the expected results from our
simulations in Section \ref{sec:expected}. Our conclusions are summarised in
Section \ref{sec:conclusions}.

Throughout this work we assume a concordance cosmology of $H_0 =
72~\rm{km}~s^{-1}\rm{Mpc}^{-1}$, $\Omega_{\rm m}$ = 0.3, and $\Omega_{\Lambda}$ =0.7, unless otherwise explicitly stated.

\bigskip

\section{CONFUSION DUE TO EXTRAGALACTIC SOURCES}\label{sec:confused}


The galaxy confusion limit is defined as the threshold of the fluctuations in
the background sky brightness below which sources cannot be discretely detected
in the telescope beam $\theta$ $\sim \lambda/D$, where $\lambda$ is the
observation wavelength and $D$ is the telescope diameter. Thus, the fluctuation
noise arises from the same origin as the galaxies that one is aiming to detect.
If we assume galaxies are distributed as a power law in flux, $S$, down to some
detection limit $S_{\rm lim}$,
\begin{equation}
 N(>S)= N_{\rm lim} \left({S\over{S_{\rm lim}}}\right)^{-\alpha},
 \label{counts}
\end{equation}
where  $N(>S)$ is the number of sources per unit solid angle with flux greater
than $S$, $\alpha$ is the slope of the integral source counts (where
$\alpha=1.5$ for a Euclidean Universe) and $N_{\rm lim}$ is the number of
sources brighter than $S_{\rm lim}$ per unit solid angle. Assuming that the
counts flatten at some faint flux, $S_{0}$, i.e. $\alpha(S_{0})=0$, then the
intensity of the background (in Jy/sr) up to some maximum flux, $S_{\rm max}$,
corresponding to these sources is given by,
\begin{equation}
  I=\!\!\int_{S_{0}}^{S_{\rm max}}\!\!\ S {dN\over{dS}} dS.
\label{intensity}
\end{equation}
The fluctuations contributed by sources below the detection limit $S_{\rm lim}$
is given by the second moment of the differential source counts $dN/dS$,
$\sigma_{\mathrm{sc}}$,
\begin{equation}
  \sigma_{\mathrm{sc}}^{2}=\!\!\int_{S_{0}}^{S_{\rm lim}}\!\!\  S^{2} {dN\over{dS}} dS.
  \label{fluctuations}
\end{equation}
Assuming the power law distribution of sources given in equation
(\ref{counts}), equation (\ref{fluctuations}) can be evaluated to give
\begin{equation}
  \sigma_{\mathrm{sc}}^{2}= N_{\rm lim}~S_{\rm lim}^{2}~{\alpha\over{2-\alpha}}
    \left[1-\left({S_{0}\over{S_{\rm lim}}}\right)^{2-\alpha}\right].
\end{equation}
For the Euclidean case, the dominant sources contributing to the background
intensity are those just below the detection limit $S_{\rm
lim}$~\cite{matsuhara00,lagache00b}. However, the strong evolution detected in
the galaxy population steepens the source counts and produces super Euclidean
slopes ($\alpha >$ 1.5) and the sources around the detection limit also
contribute significantly to the fluctuations in the background intensity.

Rigorous theoretical definitions of confusion have been presented by Scheuer
\shortcite{scheuer57} and Condon \shortcite{condon74}. Hogg \shortcite{hogg01}
has highlighted more practical aspects of galaxy confusion noise. An analytical
derivation broadly following Franceschini et al. \shortcite{franceschini89} is
given below. Note that the clustering of sources will complicate the confusion
noise [e.g., as in the case of radio sources, Condon \shortcite{condon74}]
although here, for clarity, we do not treat this effect. Some authors
\cite{franceschini89,take04,negre04} have investigated the effect of the
clustering of sources on the confusion limit. They found that clustering
possibly increases the level of the source confusion by 10\% for \textit{AKARI}
mission \cite{take04} and by 10 $\sim$ 15\% for \textit{Spitzer} and
\textit{Herschel} missions \cite{negre04} in the far-IR range.

Assuming that the sources are distributed randomly over the sky described by a
power law form $N(S) \propto S^{-\alpha}$ and a corresponding differential
distribution given by $dN/dS = kS^{-\gamma}$ where $\gamma = \alpha -1$, the
detector response to a source of flux $S$, at a position $(\theta,\phi)$ from
the axis of a detector beam of profile (point spread function) $f(\theta,\phi)$
is given by $x=Sf(\theta,\phi)$. Hence the mean number of responses with
amplitudes between $x$, $x+dx$ in a solid angle $d\Omega$ is given by;
\begin{equation}
  R(x)=\!\!\int_{\Omega_{b}}\!\!\ {n(x/f(\theta,\phi))\over{f(\theta,\phi)}} d\Omega,
\label{responses1}
\end{equation}
where $\Omega_{b}$ is the solid angle of the beam in steradians. Note that for
the power law distribution of sources discussed above, equation
(\ref{responses1}) can be rewritten as,
\begin{equation}
  R(x)= k x^{-\gamma} \!\!\int_{\Omega_{b}}\!\!\   f(\theta,\phi)^{\gamma-1} d\Omega = k x^{-\gamma} \Omega_{e},
\label{responses2}
\end{equation}
where $\Omega_{e}=\!\!\int\!\!\ f(\theta)^{\gamma -1} d\Omega$ is the effective
beam size~\cite{condon74}. Taking the second moment of the $R(x)$ distribution
(the variance) gives the fluctuation of the response, $ \sigma_{\mathrm{sc}}$ ;
\begin{equation}
  \sigma_{\mathrm{sc}}^{2}=\!\!\int_{0}^{x_{c}}\!\!\  x^{2}R(x) dx,
  \label{noise}
\end{equation}
where $x_{c}$ is a cut off response introduced to stop the variance from
diverging at bright source fluxes. More practically the confusion limit $x_{c}$
(corresponding to a cut off flux $S_{c}$) is set to some factor of the
confusion noise such that $x_{c}= q \sigma_{\mathrm{sc}}$, where the factor $q$
(usually chosen values between 3 and 5) limits the largest response included in
the evaluation of the confusion noise $\sigma_{\mathrm{sc}}$ [values of $\sim$
5 are assumed in the calculations of Franceschini et al.
\shortcite{franceschini91}]. The difference in assuming a cut off in the
response as opposed to a cut off in flux is that weak contributions from strong
sources are not neglected, as even a strong source far from the axis of the
beam may contribute significantly to the point spread function of the beam.

Assuming for clarity in the calculations, a circular Gaussian beam profile,
$f(\Theta)=f((\theta / \theta_{o})^{2})=e^{ -4 (\theta / \theta_{o})^{2}\ln2}$
such that $d\Theta=2 \theta ~d\theta / \theta_{o}^{2} $ where $\theta_{o}$ is
the Full Width at Half Maximum (FWHM) of the beam, integrating equation
(\ref{noise}) over the solid angle of the beam gives;
\begin{equation}
    \sigma_{{c}}^{2}(x_{c})=\pi \theta_{o}^{2}P(x_{c})
\end{equation}

or

\begin{equation}
\theta_{o}=  \sqrt{{\sigma_{c}^{2}\over{\pi P(x_{c})}}},
\end{equation}
where $P$, effectively the power in the fluctuations, is given by,
\begin{equation}
  P(x_{c})=\!\!\int_{0}^{x_{c}}\!\!\  x^{2}dx ~ n(x /f(\Theta)) e^{ 4 \Theta \ln2 }
  d\Theta.
\end{equation}
Thus, the confusion limit can be directly related to the FWHM of the
instrumental beam. For the simple power law representation of the distribution
of extragalactic sources given previously and the definitions of equations
(\ref{responses1}) and (\ref{responses2}), the confusion limit is given by,
\begin{equation}
  \sigma_{c}=\sqrt{ { k \Omega_{e}\over{3-\gamma}} } ~x_{c}^{(3-\gamma)/2}.
\end{equation}
Therefore, the confusion noise limit will be a complex function of the beam
size $\theta_{o}$, the source counts $N(S)$, the cut off in flux $S_{c}$ or
response $x_{c}$ and the factor $q$. For the assumed symmetric Gaussian beam
profile, $\sigma_{c} \propto \theta_{o}^{~2 / (\gamma-1)}$.

The confusion due to faint galaxies will be worse at longer wavelengths and
smaller telescope diameters. Since the confusion noise is related to the mean
number of responses (the source density) and the cut off response $q/x_{c}$, a
useful, practical benchmark for the confusion limit can be set by limiting the
number sources per beam before the beam becomes confused. Ideally, the
confusion limit would be determined by the flux at which the source density
becomes 1 source per beam although more realistically a limit of between
1/20-1/50 sources per beam (20-50 beams per source) is assumed [e.g., Hogg
\shortcite{hogg01}]. Note that this difference is due to the contribution from
faint sources undetected below a certain limit.

\section{THE INPUT CATALOGUES AND SIMULATED IMAGES}\label{sec:models}

Many authors
\cite{condon74,franceschini89,HB90,gautier92,herb98,kiss01,kiss03,jeong05} have
described the source confusion due to extragalactic point sources and the sky
confusion due to Galactic cirrus as separate issues since most extragalactic
surveys are limited to low background regions. However, in order to cover much
larger survey areas, we also naturally have to include the medium-to-high
background regions. In addition, since the cirrus structure covers both large
and small scales, both of these contributions should be considered together. By
using various source distributions and a realistically simulated cirrus
background, we address the effects of sky confusion and source confusion
simultaneously with a more realistic framework.

\subsection{Input Catalogues}

\begin{figure}
  \begin{center}
    \epsfxsize = 8.5cm
    \epsffile{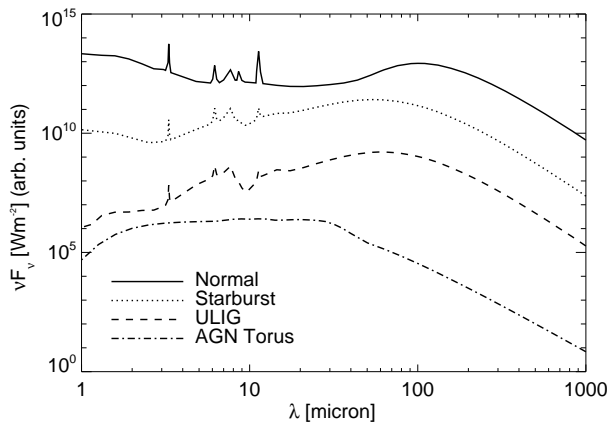}
    \end{center}
    \caption{Model input spectral energy distributions used for input
    catalogues to the simulation. A four component model comprising of
    a normal galaxy, starburst galaxy, ultraluminous galaxy and AGN dust
    torus are included. The source spectral energy distributions are based
    on the models of Efstathiou et al. (2001), Efstathiou \& Rowan-Robinson
    (2002), Rowan-Robinson (1995).}
    \label{sed}
\end{figure}

The input catalogues are prepared using the models of Pearson \shortcite{cpp01}
(hereafter CPP). CPP is a far-IR model based on the \textit{IRAS} colours and
luminosity function of galaxies [see also Pearson \& Rowan-Robinson
\shortcite{cpp96}]. The model incorporates a 4 component parameterization of
galaxy Spectral Energy Distributions (SED) segregated by \textit{IRAS} colours
\cite{mrr89}. A normal galaxy population modelled on the cool 100 $\mu$m/60
$\mu$m colours identified with infrared cirrus~\cite{low84,efs03}. A starburst
population based upon the warm 100 $\mu$m/60 $\mu$m colours of \textit{IRAS}
galaxies with the archetypical starburst galaxy M82 as a template
SED~\cite{efs00a}. An ultraluminous galaxy population~\cite{sand96}
representing the high luminosity tail of the \textit{IRAS} starburst galaxy
population and representative of the archetypical ULIG ARP220~\cite{efs00a}. An
AGN (Seyfert 1 $\&$ Seyfert 2) population modelled on a 3-30 $\mu$m dust torus
component~\cite{mrr95} defined by hot 25 $\mu$m/60 $\mu$m colours. Though
recent \textit{Spitzer} observations showed silicate dust emissions in luminous
and low-luminosity AGN \cite{hao05,sieben05,sturm05}, we do not include it. The
input SEDs are shown in Figure \ref{sed}. These SEDs have been shown to provide
good fits to the SEDs and colours of sources in the \textit{ISO}-ELAIS survey
at 6.7, 15, 90 and 170 $\mu$m \cite{mrr04} and also recently for the first
results from the \textit{Spitzer} SWIRE Legacy survey \cite{mrr05}.

To produce the input source distributions, we calculate the total number of
sources per unit solid angle at observation wavelength, $\lambda_{o}$, down to
some flux limit $S_{\lambda o}$;
\begin{equation}
   N(S_{\lambda o})=\!\!\int_{0}^{\infty}\!\! \ \!\!\int_{0}^{z(L,S)}\!\!
   \phi (L/f(z)) {dV(z)\over {dz}}~ e(z)~ d\lg L~dz,
\label{integralcount}
\end{equation}
where $f(z)$ and $e(z)$ are evolutionary functions in the luminosity and
density of the source population respectively. The integration is made over the
luminosity function (number density of objects as a function of luminosity),
$\phi (L)$ and the cosmological volume $V$, enclosed inside a limiting redshift
$z(L,S)$ defined as the redshift at which a source of luminosity, $L$, falls
below the sensitivity, $S(\lambda_{o})$.

Luminosity functions, $\phi (L)$, are determined from the All Sky \textit{IRAS}
PSCz catalogue at 60 $\mu$m by Saunders et al. \shortcite{saunders00} for
starburst and normal galaxies following colour criteria akin to those of
Rowan-Robinson $\&$ Crawford \shortcite{mrr89}. Similarly, the {\it hot} AGN
population is well represented by the 12 $\mu$m sample of Rush et al.
\shortcite{rush93} using the luminosity function of Lawrence et al.
\shortcite{lawrence86}. In addition to the above, following CPP, we utilize a
log exponential luminosity function, defined at 60 $\mu$m, to represent the
ULIG population, [referred to as the {\it Burst} model]  which was originally
implemented to address the paradigms of the strong evolution in the galaxy
source counts observed with the \textit{ISO} in the mid to far-IR, at sub-mm
wavelengths with the Submillimetre Common User Bolometer Array (SCUBA) and the
detection of the CIRB at $\sim 170\mu$m
\cite{oliver97,smail97,kawara98,hugh98,flores99,altieri99,aussel99,gruppioni99,puget99,efs00b,biviano00,elbaz00,serjeant00,lagache00a,matsuhara00,scott02}.
This model was found to provide a good fit to both the number counts and
redshift distributions of galaxies from sub-mm to near infrared wavelengths as
well as the cosmic infrared background. To shift the luminosity function from
the wavelength at which the luminosity function is defined, $\lambda _{\rm LF}
$, to the observation wavelength, $\lambda _{\rm o} $, the ratio $L(\lambda
_{\rm o})/L(\lambda _{\rm LF} )$ is obtained via the model template spectra.

We prepare 3 types of input catalogue and produce many versions of each
catalogue to reduce statistical errors.

\begin{enumerate}
  \item {\it No-Evolution Model} - No evolution is assumed for all galaxy
components.
  \item {\it Luminosity Evolution Model} - Luminosity evolution is included
following Pearson $\&$ Rowan-Robinson (1996). Only evolution in the luminosity
of the source population is assumed in this scenario. This luminosity evolution
follows a parametric form as a function of redshift of $f(z)=(1+z)^{k}
\label{levol}$, where $k$ is independently defined for each galaxy type to
produce the best fits to the far-IR counts.
  \item {\it Burst Evolution Model} -  Luminosity and density evolution is
included following the CPP model. The assumed evolution in both luminosity and
density incorporates a burst in a specific redshift range and follows a
parametric form as a function of redshift of $f(z)~ \rm{and}~ e(z) = 1+k\exp
\left[-{(z-z_{p})^{2}/{2\omega ^2}}\right]$, where the parameters $k$, $z_{p}$,
$\omega$ are independently defined for the luminosity and density evolution and
galaxy type to produce the best fits to the far-IR counts.
\end{enumerate}

\begin{figure*}
  \begin{center}
    \epsfxsize = 16cm
    \epsffile{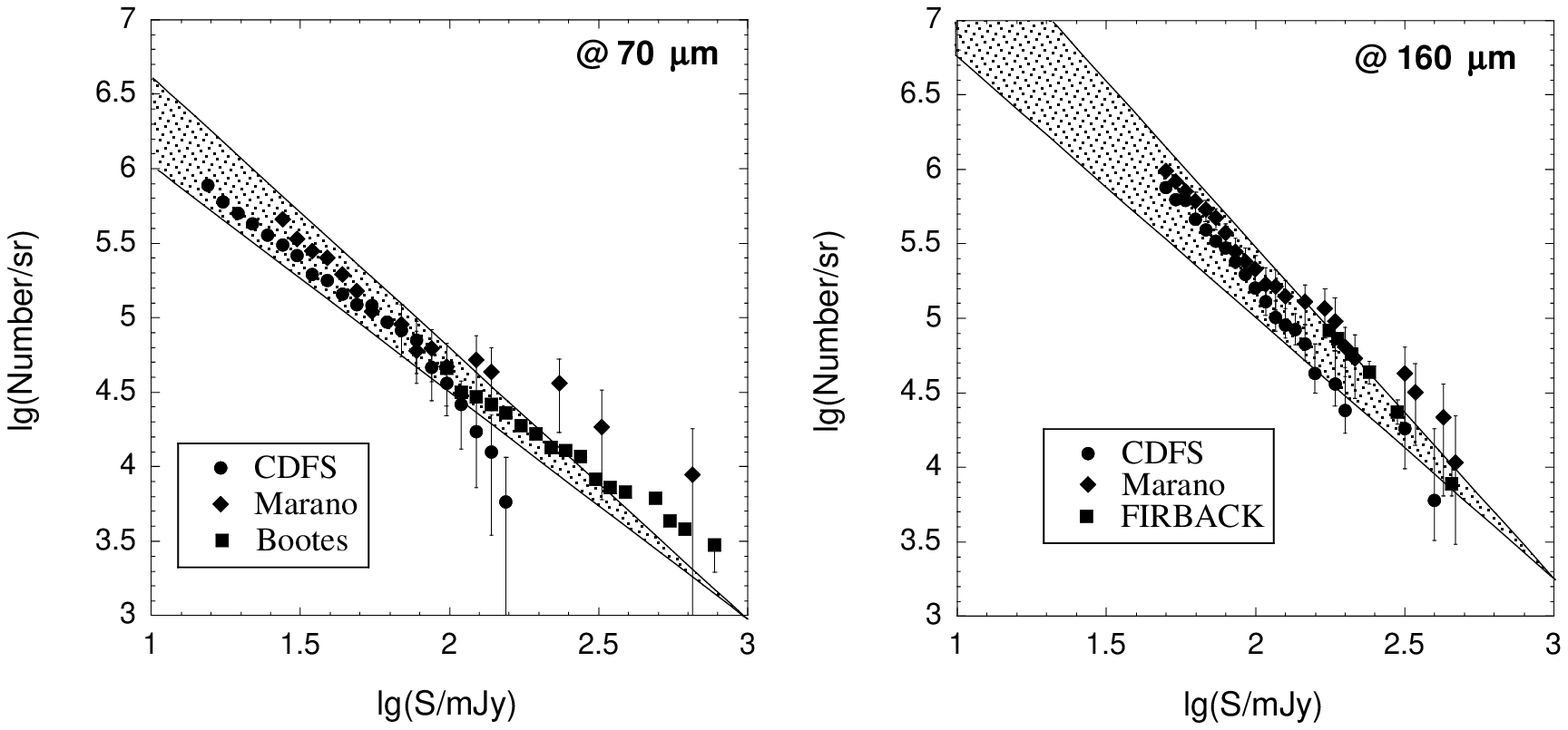}
    \end{center}
   \caption{Comparison between source count models and observations for SW
   (left) and LW (right) band. All observations except for FIRBACK field
   (Dole et al. 2001) were obtained from Multiband Imaging Photometer (MIPS)
   for \textit{Spitzer} (Rieke et al. 2004) in the Chandra Deep Field South
   (CDF-S), the Bo$\ddot{\rm o}$tes field and Marano field (Dole et al. 2004a).
   The observational mode (scan map) provides multiple sightings of each
   source, typically 10 and 60 at 70 $\mu$m in the Bo$\ddot{\rm o}$tes and
   CDF-S, respectively. Because of the low redundancy level of the 160 $\mu$m
   data in the Bo$\ddot{\rm o}$tes field, they exclude those data. Upper and
   lower lines show our burst evolution model and luminosity evolution model,
   respectively, with the hashed area defining upper and lower bounds of far-IR
   source distributions assumed in this work.}
   \label{fig_comp_scount}
\end{figure*}

Details of the evolution included in the models can be found in CPP. Figure
\ref{fig_comp_scount} shows the evolutionary source count models for the
\textit{Spitzer} far-IR galaxy counts at 70 $\mu$m and 160 $\mu$m respectively
\cite{dole04a}. The two evolutionary models provide reasonable fits and also
cover the spread in uncertainty shown in the observations. Therefore, in this
work we consider the \textit{burst evolution} and  \textit{luminosity
evolution} models as upper and lower limits to the numbers of far-IR sources
respectively. We do not attempt to fit the mid-near-infrared \textit{Spitzer}
counts. For specific updated evolutionary models, please see Pearson
\shortcite{cpp05} and Pearson \shortcite{cpp06a}.

\subsection{Simulated Images}

Based upon the 3 types of input catalogue, we generated simulated images in
each band for various space missions. The image size for the distributed source
simulation is $\sim$ 5.3 square degrees. In order to check the effect of sky
confusion noise, we also include high resolution cirrus maps whose mean cirrus
brightness ranges from 0.3 MJy~sr$^{-1}$ to 25 MJy~sr$^{-1}$ at 100 $\mu$m as
the background by using the method described in Paper I.

\bigskip

\section{SIMULATION RESULTS}\label{sec:simulation}

As we described in Section \ref{sec:confused}, there are many definitions for
the source confusion. We compare these definitions and propose an appropriate
definition for source confusion. In Table \ref{tab_inst_para}, we list the
basic instrumental parameters of present and future IR space missions; the
aperture of the telescope, Full Width at Half Maximum (FWHM) of the beam
profile and the pixel size for each detector.

\begin {table*}
\centering
\caption {Instrumental parameters for various space missions.}
\label{tab_inst_para} \vspace{5pt}
\begin{tabular}{@{}ccrlrlrl}
\hline\vspace{-5pt} \\
& Aperture & \multicolumn{2}{c}{Central Wavelength}& \multicolumn{2}{c}{FWHM $^a$} & \multicolumn{2}{c}{Pixel size} \vspace{5pt} \\
 & (meter) & \multicolumn{2}{c}{($\mu$m)} & \multicolumn{2}{c}{(arcsec)} & \multicolumn{2}{c}{(arcsec)} \vspace{5pt} \\
Space Mission & & ~~~~~~SW & LW & SW & LW & SW & LW \vspace{5pt}
\\\hline \vspace{-10pt}
\\ \textit{Spitzer} $^b$ & 0.85 & 70 & 160 & 17 & 35 & 10 & 16 \vspace{5pt}
\\ \textit{AKARI} $^c$ & 0.69 & 75 & 140 & 23 & 44 & 27 & 44 \vspace{5pt}
\\ \textit{Herschel} $^d$ & 3.5 & 70 & 175 $^e$ &  4.3 & 9.7 & 3.2 & 6.4 \vspace{5pt}
\\ \textit{SPICA} & 3.5 & 70 & 160 &  4.3 & 9.7 & 1.8 & 3.6 \vspace{5pt}
\\ \hline
\end{tabular}

\medskip
\begin{flushleft}
{\em $^a$} FWHM of diffraction pattern. \\
{\em $^b$} Two MIPS bands for the \textit{Spitzer} mission have 3 bands with
central wavelengths of 24 $\mu$m, 70 $\mu$m and 160 $\mu$m.
We use 70 $\mu$m band as SW band and 160 $\mu$m band as LW band in this paper.\\
{\em $^c$} \textit{AKARI/FIS} (Far Infrared Surveyor) has a WIDE-S band in SW
and a WIDE-L band in LW. \\
{\em $^d$} PACS has a `blue' array in two short wavelength bands (centered at 70
$\mu$m and 110 $\mu$m) and a `red' array at longer wavelengths (centered at
175$\mu$m). In this paper, we use only  the 70 $\mu$m band of the `blue' array.\\
{\em $^{b,d}$} Since one of our motivations in this paper is to compare the
confusion limits, we use only common bands among all far-IR bands in the considered space missions. \\
{\em $^e$} Note that though the central wavelength of the `red' array in PACS is
175 $\mu$m, our estimated results are for 160 $\mu$m. Due to this wider beam,
the confusion noise is expected to increase by $\sim$ 15\% at 175 $\mu$m compared
to that at 160 $\mu$m.
\end{flushleft}
\end{table*}

\subsection{Definition by `Beams per Source'}\label{sec:sc_def_beam}

\begin{figure*}
  \begin{center}
    \epsfxsize = 8.5cm
    \epsffile{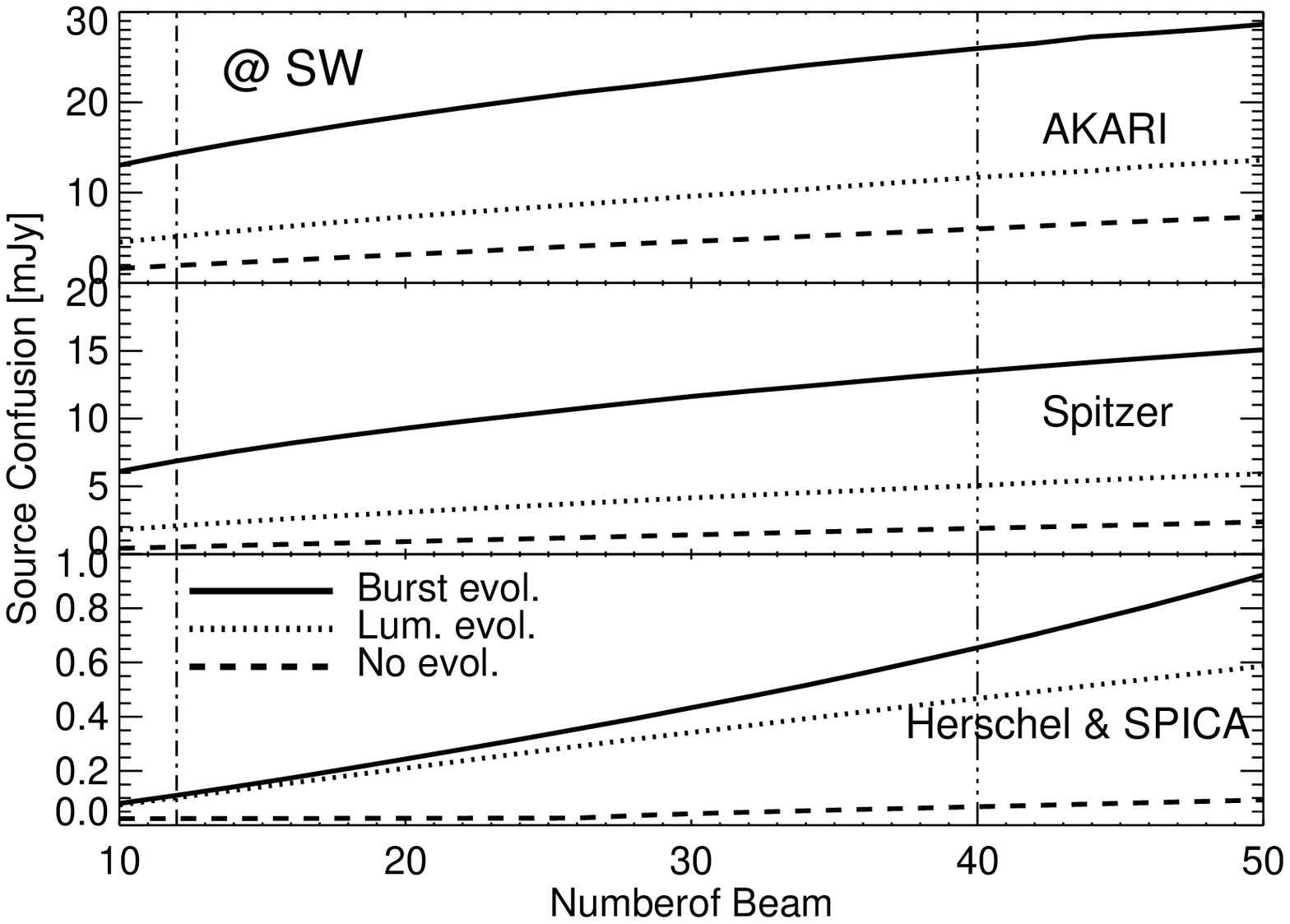}
    \epsfxsize = 8.5cm
    \epsffile{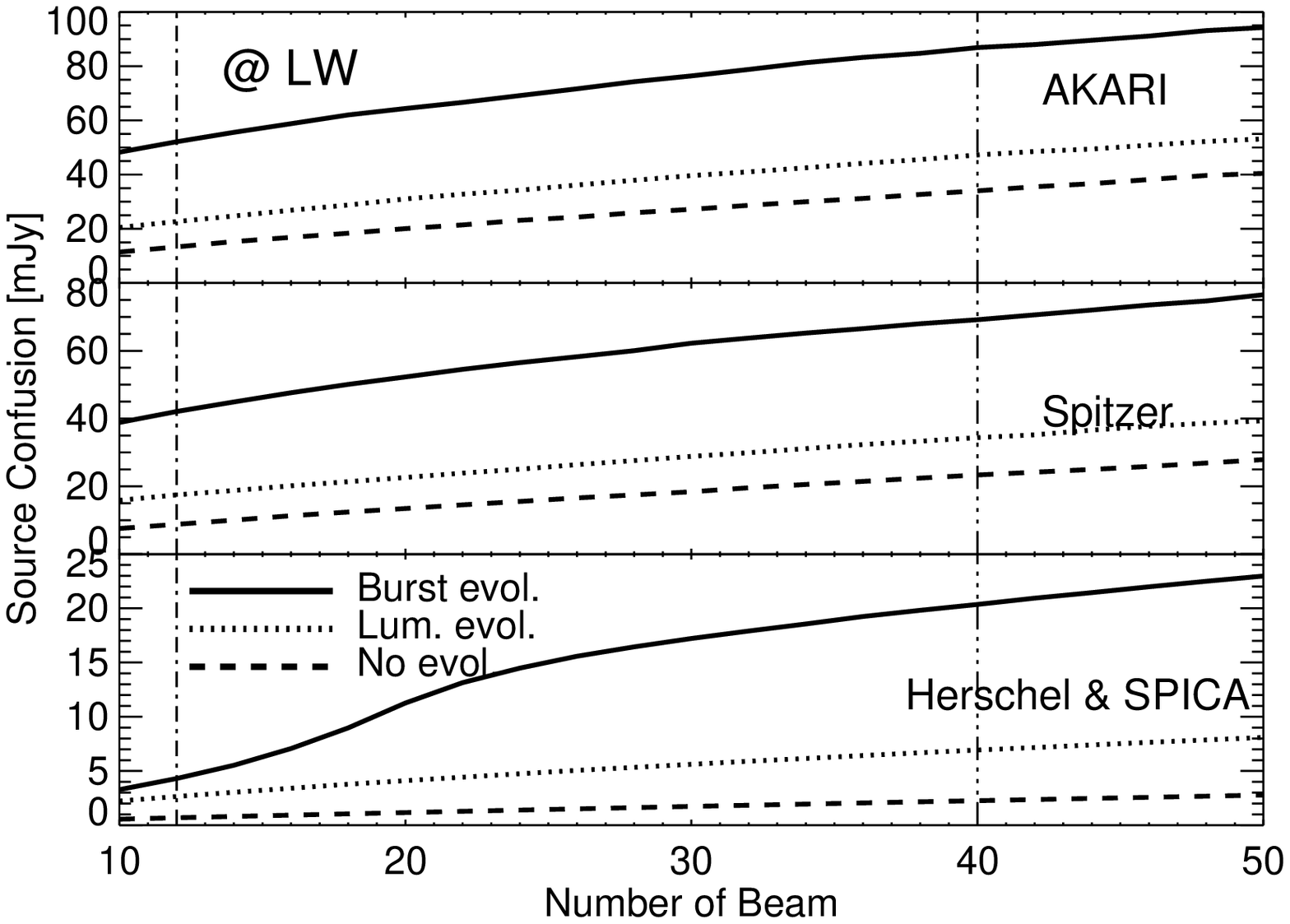}
    \end{center}
   \caption{Source confusion with the definition of beams per sources for
   \textit{AKARI}, \textit{Spitzer}, and \textit{Herschel} \& \textit{SPICA}
   missions. Each line shows the source distribution model used in this paper.
   The two vertical lines show 12 beams (left) and 40 beams (right) per source,
   respectively.}
   \label{fig_beam_sconf}
\end{figure*}

First, we estimate the source confusion from the classical definition of source
confusion, 'beams per source' (sources per beam). Though often cited in the
literature as 'sources per beam', we use the term 'beams per source' for our
definition of source confusion in order to compare the effect of source
confusion according to the different beam sizes  for each mission. We check the
source confusion by changing the number of beams per source criterion from 10
to 50. Fig. \ref{fig_beam_sconf} shows the source confusion assuming this
definition for each mission and each model. Hogg \shortcite{hogg01} showed that
30 beams per source is the minimum photometric criterion where the source
counts are steep, and suggested 50 beams per source for the definition of
source confusion. Rowan-Robinson \shortcite{mrr01} adopted 40 beams per source.
In recent papers, Dole et al. (2003, 2004b) suggested that source confusion
could be defined by a source density criteria, corresponding to $\sim$ 12 beams
per source for the \textit{Spitzer} mission. They have estimated a source
confusion limit  for the \textit{Spitzer} mission of 3.2 mJy for SW band and 40
mJy for LW band, respectively. These results are similar to our estimations
with the definition of 12 beams per source for the luminosity evolution model
in the SW band and for the burst evolution model in the LW band, respectively.
These consistences suggest that the source distribution model used for the
estimation of the  source confusion limit for the case of \textit{Spitzer}
\cite{lagache03,lagache04} also predicts that a starburst component, and  ULIG
component dominate the SW and LW bands respectively. In Table
\ref{tab_sc_beams}, we list the source confusion limits estimated by the
definition of 12 and 40 beams per source. As seen from Fig.
\ref{fig_beam_sconf}, we found that the source confusion did not increase at a
constant rate and that the slope of the source confusion slightly varies
according to the source distribution model and the resolution of the mission,
especially for burst evolution model. Therefore, we conclude that we can not
apply the same definition of 'beams per source' in a generic way for different
missions.

\begin {table*}
\centering \caption[Source confusion estimated by the definition of 12 and 40
beams per source] {{Source confusion estimated by the definition of 12 and 40
beams per source.}} \label{tab_sc_beams} \vspace{5pt}
\begin{tabular}{@{}ccccccccccccc}
\hline\vspace{-5pt} \\
& \multicolumn{4}{c}{No evolution} & \multicolumn{4}{c}{Luminosity evolution} & \multicolumn{4}{c}{Burst evolution} \vspace{5pt} \\
& \multicolumn{4}{c}{(mJy)} & \multicolumn{4}{c}{(mJy)} & \multicolumn{4}{c}{(mJy)} \vspace{5pt} \\
& \multicolumn{2}{c}{SW} & \multicolumn{2}{c}{LW} & \multicolumn{2}{c}{SW} &
\multicolumn{2}{c}{LW} & \multicolumn{2}{c}{SW} & \multicolumn{2}{c}{LW}
\vspace{5pt} \\ Space Mission & 12b~$^a$ & 40b~$^b$ & 12b & 40b & 12b & 40b &
12b & 40b & 12b & 40b & 12b & 40b \vspace{5pt}
\\\hline \vspace{-10pt}
\\ \textit{Spitzer} & 0.53 & 1.9 & 8.8 & 23 & 2.1 & 5.1 & 17 & 34 & 6.9 & 13 & 42 & 69 \vspace{5pt}
\\ \textit{AKARI} $^c$ & 1.9 & 6.0 & 13 & 34 & 5.1 & 12 & 23 & 47 & 14 & 26 & 52 & 87 \vspace{5pt}
\\ \textit{Herschel} \& \textit{SPICA} & 0.014 & 0.066 & 0.65 & 2.1 & 0.10 & 0.47 & 2.6 & 6.8 & 0.11 & 0.65 & 4.3 & 20 \vspace{5pt}
\\ \hline
\end{tabular}

\medskip
\begin{flushleft}
{\em $^a$} source confusion defined by Dole et al. (2004b): flux corresponding to 12 beams per source. \\
{\em $^b$} source confusion defined by Rowan-Robinson (2001): flux corresponding to 40 beams per source. \\
{\em $^c$} Wide-S band for SW and Wide-L band for LW.
\end{flushleft}
\end{table*}

\subsection{Definition by Fluctuation}\label{sec:sc_def_fluc}

\begin{figure}
  \begin{center}
    \epsfxsize = 8.5cm
    \epsffile{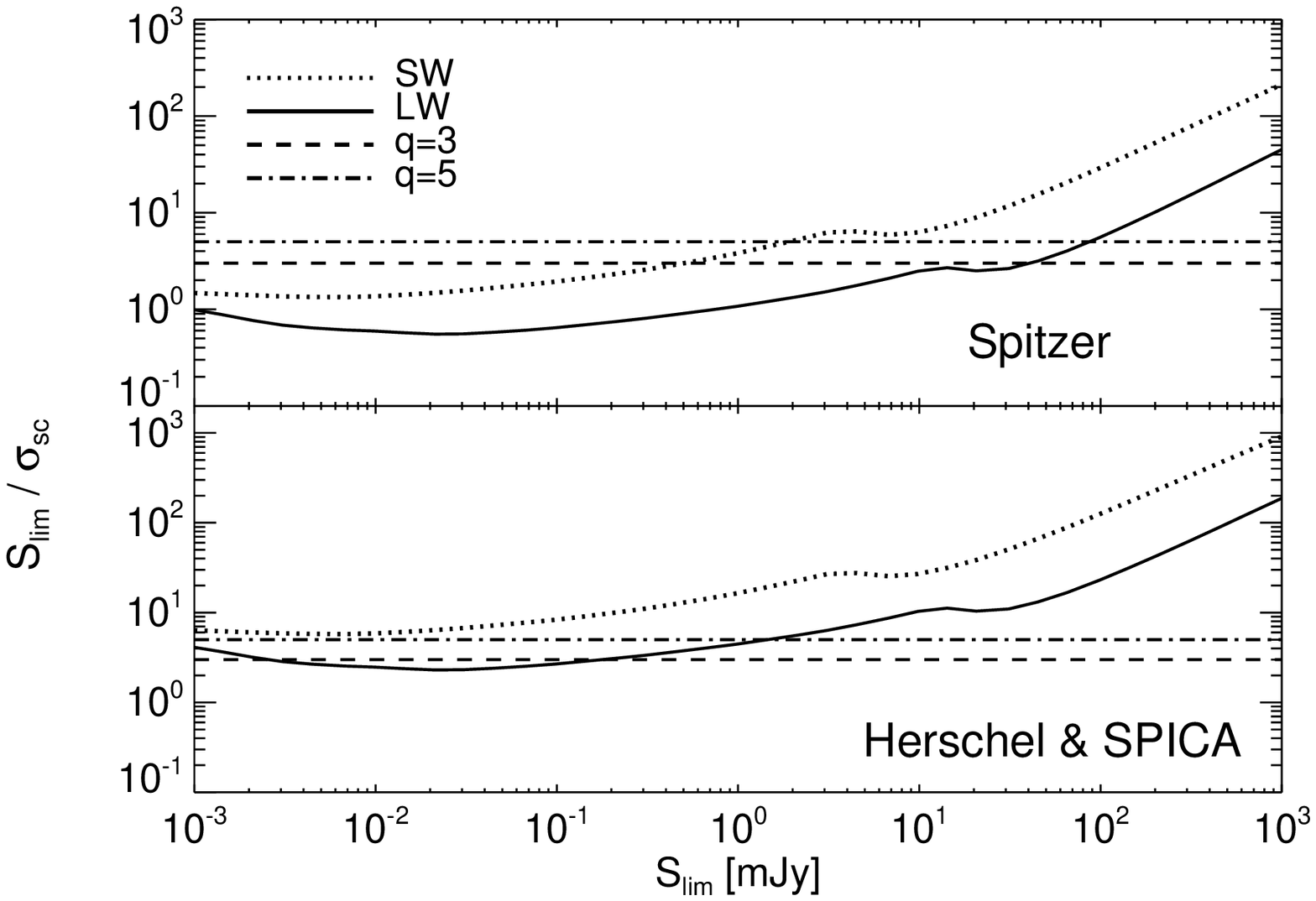}
    \end{center}
   \caption{$S_{\mathrm{lim}} / \sigma_{\mathrm{sc}}$ ratio as a function of
   $S_{\mathrm{lim}}$ for \textit{Spitzer} (upper) and \textit{Herschel}
   \& \textit{SPICA} (lower) missions in the case of a burst evolution model.
   We also plot $S_{\mathrm{lim}} / \sigma_{\mathrm{sc}}$ = 3 (dashed line) and 5
   (dashed dotted line). In the case of the \textit{Herschel} \& \textit{SPICA}
   missions in the SW band, note that $S_{\mathrm{lim}} / \sigma_{\mathrm{sc}}$ is
   always greater than 5, and the plateaus in the SW and LW band are due to
   the bump seen in the source counts.}
   \label{fig_sn_sconf}
\end{figure}

Another criterion for the quantification of source confusion can be defined by
the fluctuation from beam to beam due to the distribution of the point sources
\cite{condon74,franceschini89,vaisanen01,xu01}. Since the beam size is large
and the source counts are steep, the usual definition of `beams per source' may
not be valid in the case of far-IR photometry. For the PSF, we assume an ideal
circular aperture Airy pattern corresponding to the aperture size of the
telescopes for each mission except for \textit{AKARI} for which we use  the
theoretical PSFs estimated from the telescope design \cite{jeong03}. To
visualize the iteration procedure, we plot the ${S_{\mathrm{lim}} /
\sigma_{\mathrm{sc}}(S_{\mathrm{lim}})}$ ratio as a function of
$S_{\mathrm{lim}}$ for the \textit{Spitzer} and \textit{Herschel} \&
\textit{SPICA} missions for the case of the burst evolution model (see Figure
\ref{fig_sn_sconf}). For the SW band of the \textit{Herschel} \& \textit{SPCIA}
missions, the ${S_{\mathrm{lim}} / \sigma_{\mathrm{sc}}}$ ratio is always
greater than 5, which means that we can not obtain a solution for source
confusion, even for $q$ = 5. Dole et al. \shortcite{dole03} estimated the
source confusion with their best estimator for the \textit{Spitzer} mission.
With optimized $q$ parameters of 3.8 in SW band and 6.8 in LW band, they
obtained source confusion limits of 3.2 mJy and 36 mJy at 70 $\mu$m and 160
$\mu$m, respectively. Using our source count models and the same $q$
parameters, we obtained the source confusion limits of 3.7 mJy and 21 mJy for
the luminosity evolution model and 12 mJy and 60 mJy for the burst evolution
model at the same wavelengths. As we showed in Section \ref{sec:sc_def_beam},
though the source density of the source count model used in Dole et al.
\shortcite{dole03} is similar to the luminosity evolution model in the SW band
and the burst evolution model in the LW band, we find that the fluctuation due
to distributed sources in the burst evolution model is much stronger than that
of the model of Dole et al. \shortcite{dole03}. This amount of fluctuation may
degrade the final detection limits.

In Table \ref{tab_sc_fluc}, we list the source confusion limits estimated from
the definition by fluctuation for the cases of $q$ = 3 and $q$ = 5. For $q$ =
5, there are no solutions for the SW band of \textit{Herschel} \&
\textit{SPICA} missions. Even in the LW band, we could not find a reasonable
solution since the estimated source confusion for the two evolutionary models
gave identical results. However, when we attempted photometry on the simulated
images including the point sources for each source count model, we encountered
some limits to the source detection. Therefore, we conclude that we can not
apply a constant $q$ to the estimation of source confusion generically for all
cases.

\begin {table*}
\centering \caption{Source confusion estimated by the definition of fluctuation
with $q$ = 3 and $q$ = 5.} \label{tab_sc_fluc} \vspace{5pt}
\begin{tabular}{@{}ccccccccccccc}
\hline\vspace{-5pt} \\
& \multicolumn{4}{c}{No evolution} & \multicolumn{4}{c}{Luminosity evolution} & \multicolumn{4}{c}{Burst evolution} \vspace{5pt} \\
& \multicolumn{4}{c}{(mJy)} & \multicolumn{4}{c}{(mJy)} & \multicolumn{4}{c}{(mJy)} \vspace{5pt} \\
& \multicolumn{2}{c}{SW} & \multicolumn{2}{c}{LW} & \multicolumn{2}{c}{SW} &
\multicolumn{2}{c}{LW} & \multicolumn{2}{c}{SW} & \multicolumn{2}{c}{LW}
\vspace{5pt} \\ Space Mission & $q$ = 3 & $q$ = 5 & $q$ = 3 & $q$ = 5 & $q$ = 3
& $q$ = 5 & $q$ = 3 & $q$ = 5 & $q$ = 3 & $q$ = 5 & $q$ = 3 & $q$ = 5
\vspace{5pt}
\\\hline \vspace{-10pt}
\\ \textit{Spitzer} & 0.065 & 0.28 & 3.8 & 10 & 0.51 & 1.9 & 14 & 33 & 0.51 & 1.9 & 41 & 87 \vspace{5pt}
\\ \textit{AKARI} & 0.36 & 1.4 & 5.6 & 15 & 2.4 & 6.8 & 20 & 43 & 2.4 & 23 & 59 & 115 \vspace{5pt}
\\ \textit{Herschel} \& \textit{SPICA} & $-$~$^a$ & 0.002 & 0.042 & 0.34 & $-$ & 0.006 & 0.18 & 1.5 & $-$ & $-$ & 0.18 & 1.5 \vspace{5pt}
\\ \hline
\end{tabular}

\begin{flushleft}
{\em $^a$} no solution for this case.
\end{flushleft}
\end{table*}

\subsection{Composite Definition by Fluctuation and Photometry}

In an attempt to define the source confusion for all cases, we implement a
composite definition by fluctuation and photometry. As a first step, we attempt
photometry on the simulated images created from each source count model. The
definition of completeness and reliability for quantifying the source detection
efficiency is widely used in photometry. The `completeness' is defined as the
fraction of detected sources to the original input catalogue sources and the
`reliability' as the fraction of real sources to all detected sources.For our
photometry, we removed the false or spurious detections by comparing the
positions of detected sources with those in the input catalogue. The real
sources as defined in the reliability, refers to sources of which the measured
flux agrees with the input flux to within a 20\% error. Note that the
reliability at a given flux range is not always higher than the completeness.
An excess of sources near the detection limit, or more likely an overestimation
of the flux of sources at or near the detection limit can be caused by a step
effect where the underlying, unresolved sources are entering the PSF, affecting
the sky-subtraction. We carried out aperture photometry on the simulated images
using the SExtractor software \textit{v}2.2.2~\cite{bert96}. The most
influential parameters are the size of background mesh and the threshold for
the source detection for the aperture photometry. Since the size of background
mesh is related to the range of scales of the background fluctuation, as we set
a smaller mesh size, we can detect smaller fluctuation. Therefore, if we use
smaller size of background mesh and thresholds, we obtained many more spurious
sources. We set the size of the background mesh to $\sim$ 2 times the measuring
aperture, and the detection threshold as 4, which is optimized for better
reliability of the detected sources thus reducing the false detection rate.

As discussed in Section \ref{sec:sc_def_fluc}, we could not use a constant
value of $q$ for the estimation of source confusion. Current and future space
missions, will detect much fainter sources to higher sensitivity, which means
that we will observe high source densities to extremely faint detectable flux
levels. Therefore, a significant factor contributing to the source detection
comes from both the faint sources below the detection limit and the high source
density above or around the detection limit. With this assumption, we include
the contributed  fluctuations from the sources above the detection limit as
well. In order to find the limiting flux affecting the source detection, we use
a photometric method on the simulated images. We set the limiting flux
$S_{\mathrm{lim}}$ in equation \ref{fluctuations} to be the flux that the
completeness reaches $\sim$ 80\% where the  completeness level abruptly falls
off. We do not use the reliability criterion in this definition since the
reliability can be improved by using optimized photometric methods. We also
assume that the sources above this flux level do not contribute significantly
to the source confusion. We obtain the final source confusion from a 4$\sigma$
fluctuation in order to compare with the threshold used in the photometry. In
Table \ref{tab_sc_phot}, we list the source confusion limits from our best
estimators.

\begin {table}
\centering \caption[Source confusion estimated by the definition of the
composition of fluctuation and photometry]{{Source confusion estimated by the
composite definition of both fluctuation and photometry.}} \label{tab_sc_phot}
\vspace{5pt}
\begin{tabular}{@{}crlrlrl}
\hline\vspace{-5pt} \\
& \multicolumn{2}{c}{No evol.} & \multicolumn{2}{c}{Lum. evol.} & \multicolumn{2}{c}{Burst evol.} \vspace{5pt} \\
& \multicolumn{2}{c}{(mJy)} & \multicolumn{2}{c}{(mJy)} & \multicolumn{2}{c}{(mJy)} \vspace{5pt} \\
Space Mission & SW & LW & SW & LW & SW & LW \vspace{5pt}
\\\hline
\vspace{-10pt}
\\ \textit{Spitzer} & 0.48 & 9.4 & 2.0 & 23 & 6.5 & 66 \vspace{5pt}
\\ \textit{AKARI} & 2.0 & 14 & 6.3 & 33 & 17 & 87 \vspace{5pt}
\\ \textit{Herschel} \& \textit{SPICA} & 0.0095 & 0.63 & 0.077 & 2.2 & 0.10 & 4.4 \vspace{5pt}
\\ \hline
\end{tabular}
\end{table}

Though we could not obtain the results for the SW band of \textit{Herschel} \&
\textit{SPICA} missions with the original definition of the fluctuation with constant
$q$, we could obtain the results by our new best estimator. We also find that the
source confusion limits by our definition are mostly consistent with a
completeness of $\sim$ 65\% for all missions. Therefore, we conclude that our
definition can explain the behaviour of source confusion well in the far-IR range,
regardless of individual space mission characteristics.

\subsection{Predicted Confusion Limits for Current and Future Missions}
\label{sec:conf_limits}

\begin{figure}
  \begin{center}
    \epsfxsize = 7.5cm
    \epsffile{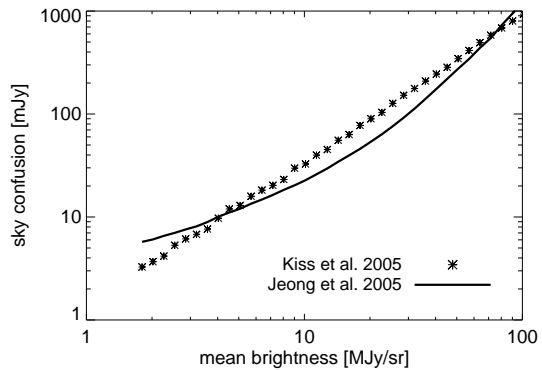}
    \end{center}
   \caption{Comparison of sky confusion between Kiss et al. (2005) and Jeong et al.
   (2005) for the \textit{AKARI} mission at 170 $\mu$m. Two results shows significant
   differences in the range of low and medium cirrus brightness.}
   \label{fig_comp_kiss}
\end{figure}

In our study of sky confusion in Paper I, we predicted the confusion limits
considering both sky confusion due to cirrus structure and source confusion for
each mission from a rather simple (power law) source distribution models and
plotted the sky confusion levels for an assumed range of average cirrus
brightness  $\langle B_{\lambda} \rangle$ in Table \ref{tab_sc_cirrus} (see
details in Paper I). Kiss et al. \shortcite{kiss05} have also estimated the
all-sky cirrus confusion for various IR missions. Fig. \ref{fig_comp_kiss}
shows a comparison of sky confusion between Kiss et al. \shortcite{kiss05} and
Jeong et al. \shortcite{jeong05} for the \textit{AKARI} mission at 170 $\mu$m.
Since the Kiss et al. (2005) model assumes a linear relationship between the
cirrus brightness and sky confusion, it often gives different results in low to
medium cirrus brightness.

%
\begin {table}
\centering \caption[Estimated confusion limits due to Galactic cirrus for each
mission]{{Estimated confusion limits due to Galactic cirrus for each mission.}}
\label{tab_sc_cirrus} \vspace{8pt}
\begin{tabular}{@{}crlrlrl}
\hline\vspace{-5pt} \\
& \multicolumn{6}{c}{Sky Confusion Limits (mJy) for $\langle B_{\lambda} \rangle$$^a$} \vspace{5pt} \\
& \multicolumn{2}{c}{0.5 MJy/sr} & \multicolumn{2}{c}{5 MJy/sr} & \multicolumn{2}{c}{15 MJy/sr} \vspace{5pt} \\
Space Mission & SW & LW & ~~SW & LW & ~~SW & LW \vspace{5pt}
\\\hline \vspace{-10pt}
\\ \textit{Spitzer} & 0.10 & 7.1 & 1.2 & 10 & 11 & 19 \vspace{5pt}
\\ \textit{AKARI} & 0.73 & 9.4 & 5.4 & 18 & 27 & 42 \vspace{5pt}
\\ \textit{Herschel} \& \textit{SPICA} & 0.003 & 0.16 & 0.027 & 0.21 & 0.31 & 0.37 \vspace{5pt}
\\ \hline
\end{tabular}

\begin{flushleft}
{\em $^a$} mean brightness of cirrus emission
\end{flushleft}
\end{table}

\begin{figure*}
  \begin{center}
    \epsfxsize = 8.5cm
    \epsffile{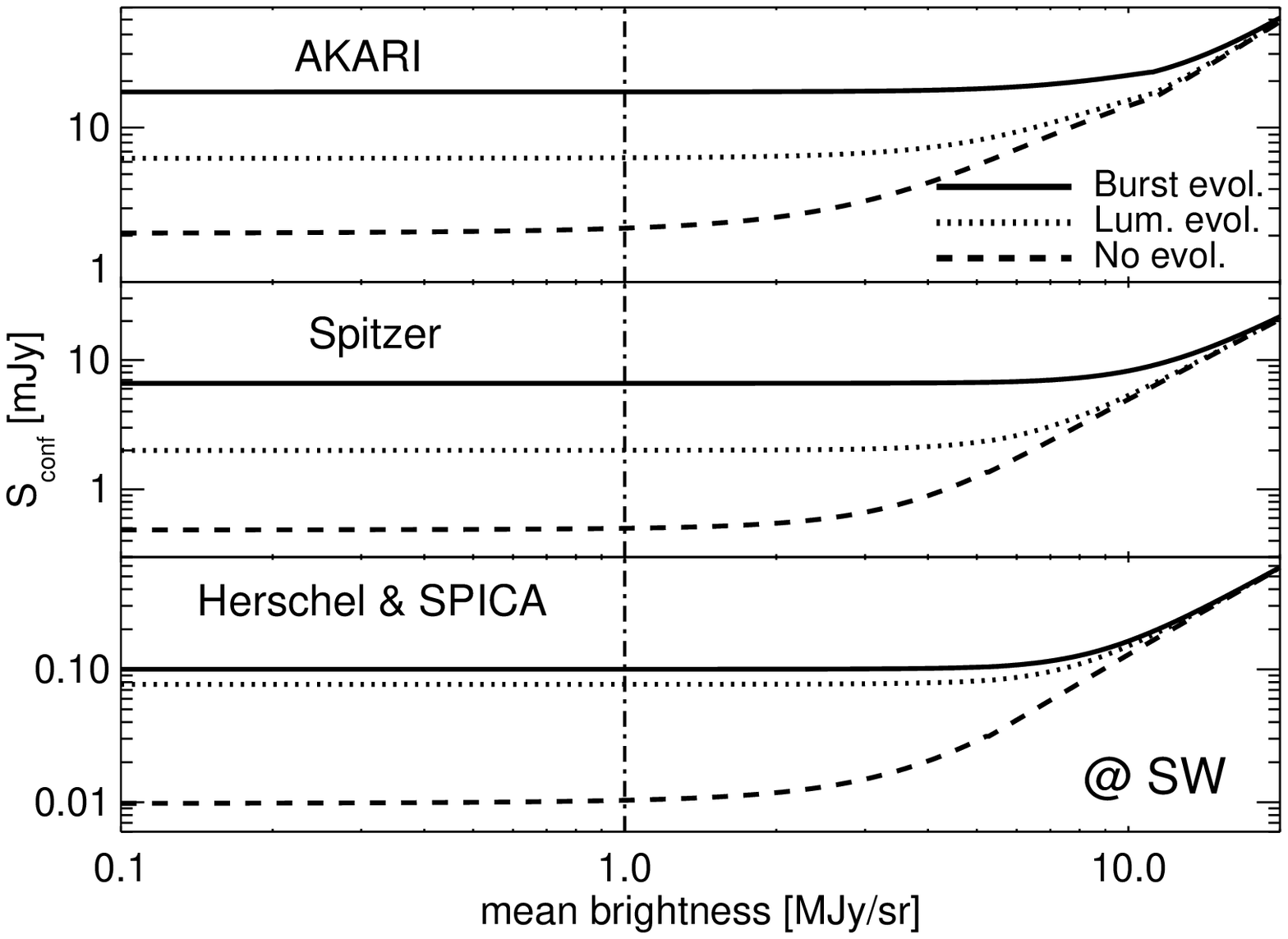}
    \epsfxsize = 8.5cm
    \epsffile{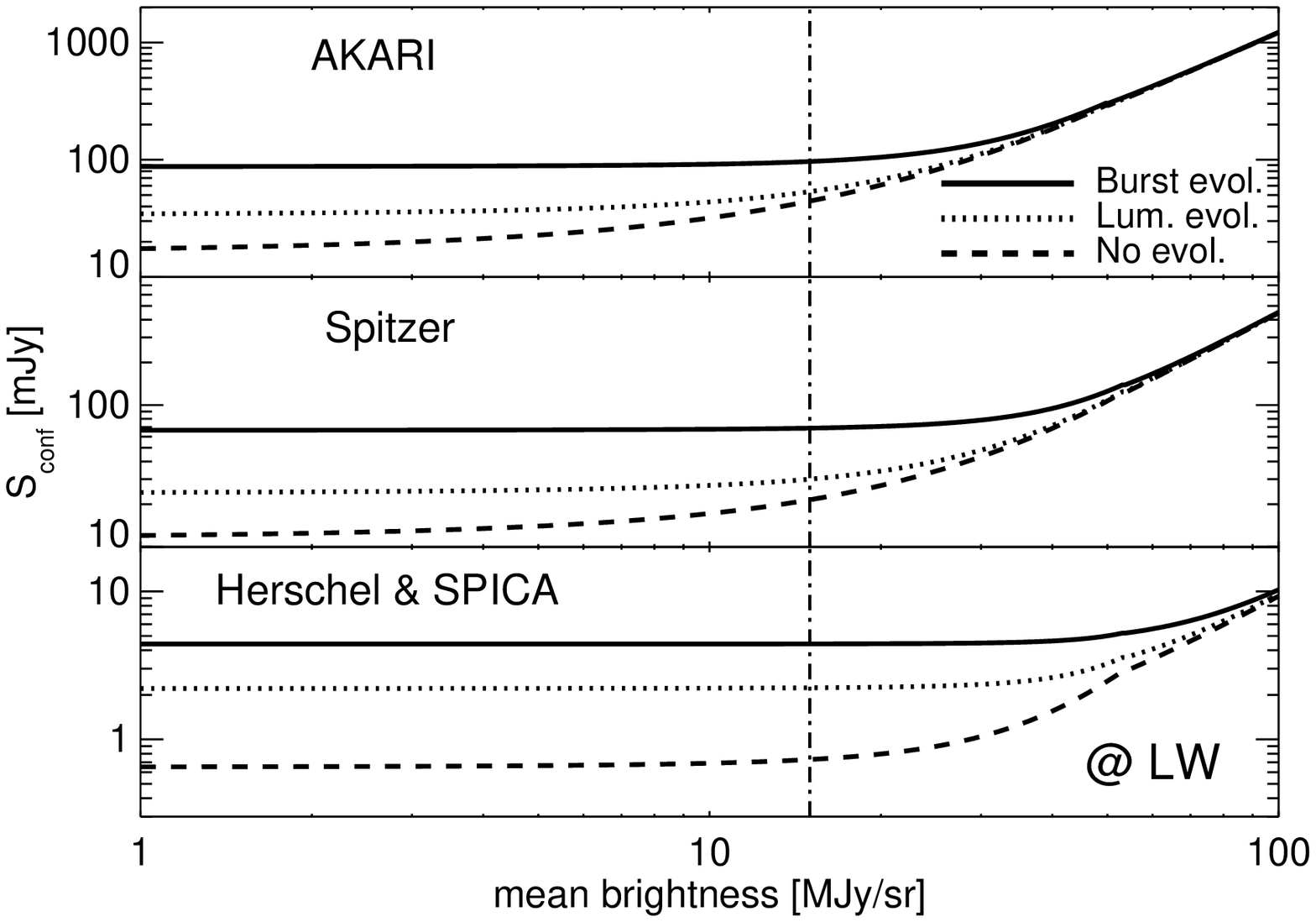}
    \end{center}
   \caption{Confusion limits considering both sky confusion and source confusion.
   The two vertical lines show the mean cirrus brightness 1.0 MJy~sr$^{-1}$
   for SW band (left) and 15 MJy~sr$^{-1}$ LW band (right), respectively. In
   dark cirrus regions, source confusion is expected to dominate. Note that
   sky confusion increases as the mean cirrus brightness becomes larger and
   the source confusion has a constant value irrespective of the mean cirrus
   brightness.}
   \label{fig_conf_all}
\end{figure*}

Since, in general, the cirrus fluctuations are not represented by Gaussian
noise, we can not directly sum the two noise contributions (sky confusion and
source confusion) for the estimation of the final confusion noise. However, in
a recent paper, H$\acute{\rm e}$raudeau et al. \shortcite{herau04} found that
the distribution of cirrus fluctuation is near-Gaussian, at least at the
ISOPHOT 90 and 170 $\mu$m spatial frequencies. Thus, we use equation
\ref{eq_tot_no} to estimate final confusion limits $S_{\rm conf}$ in this paper
by summation of the two confusion noise components:
\begin{equation}
S_{\rm conf} = \sqrt{S_{\rm cc}^2 + S_{\rm sc}^2}, \label{eq_tot_no}
\end{equation}
where $S_{\rm cc}$ is the sky confusion limit due to cirrus structure and
$S_{\rm sc}$ is the source confusion limit. Fig. \ref{fig_conf_all} shows the
final confusion limits due to both sky and source confusion for each mission
and each evolutionary model. The sky confusion will be negligible for the
\textit{Herschel} and \textit{SPICA} missions even in regions of high cirrus
brightness and the dominant confusion contribution is predicted to be the
source confusion in high Galactic latitude regions for the purpose of
cosmological studies. However, care must be taken when considering the
\textit{Spitzer} and \textit{AKARI} missions in the mean cirrus brightness
range $\langle B_{\lambda}\rangle$ $>$ 20 MJy~sr$^{-1}$ for the LW band. In order
to check the effect caused by the combination of both sky and source confusion,
we attempt photometry on the simulated images including various cirrus
background levels and source count models and compare them with the photometric
results considering only source confusion. These results are presented in
Appendix \ref{app:phot_results}.

We summarize the confusion limits including both sky and source confusion for
each mission in Table \ref{tab_final_sc} for the low cirrus background regions
dominated by source confusion. The mean brightness used in the estimation of
sky confusion is 0.5 MJy~sr$^{-1}$ and 1.0 MJy~sr$^{-1}$ for the SW and LW bands,
respectively. The sensitivity for the \textit{AKARI} mission is estimated based
upon the recent laboratory experiments, and various detector characteristics
and observational environments which may affect the final sensitivity (Jeong et
al. 2003, 2004; Matsuura et al. 2002; Shirahata et al. 2004). Though the single
scan sensitivity in survey mode of the \textit{AKARI} mission is much lower
than the confusion limit, the sensitivity at higher ecliptic latitudes where
many scans overlap  should approach the confusion limit. For the
\textit{Herschel} mission, since the sensitivity is 3 mJy, we expect that most
observations in the SW band will be dominated by instrumental noise and those
in the LW band will be near the confusion limit. However, the \textit{SPICA}
mission will achieve the confusion limit in both bands since it will have a
large aperture telescope cooled to very low temperatures \cite{naka04}.

\begin {table*}
\centering \caption[Final confusion limit] {{Final confusion limits considering
both source confusion and sky confusion.}} \label{tab_final_sc} \vspace{5pt}
\begin{tabular}{@{}cccrlrlrl}
\hline\vspace{-5pt} \\
& \multicolumn{2}{c}{Sensitivity} & \multicolumn{2}{c}{No evolution} & \multicolumn{2}{c}{Luminosity evolution} & \multicolumn{2}{c}{Burst evolution} \vspace{5pt} \\
& \multicolumn{2}{c}{(mJy)} & \multicolumn{2}{c}{(mJy)} & \multicolumn{2}{c}{(mJy)} & \multicolumn{2}{c}{(mJy)} \vspace{5pt} \\
Space Mission & SW & LW & SW & LW & ~~~~~~~~SW & LW & ~~~~SW & LW \vspace{5pt}
\\\hline \vspace{-10pt}
\\ \textit{Spitzer} & 6 & 15 & 0.49 & 12 & 2.0 & 25 & 6.6 & 67 \vspace{5pt}
\\ \textit{AKARI} $^a$ & 12 (200) & 12 (400) & 2.2 & 18 & 6.4 & 34 & 17 & 88 \vspace{5pt}
\\ \textit{SPICA} $^b$ & 3 & 3 & 0.010 & 0.66 & 0.077 & 2.3 & 0.10 & 4.4 \vspace{5pt}
\\ \textit{Herschel} $^c$ & 3 & 3 & 0.010 & 0.73 & 0.077 & 2.6 & 0.10 & 5.1 \vspace{5pt}
\\ \hline
\end{tabular}

\medskip
\begin{flushleft}
{\em $^a$} 5$\sigma$ sensitivity for slow scan mode (scan speed: 8 arcsec/sec,
reset: 1 sec) and survey mode (parenthesized values, 1 scan). The sensitivity
of the survey mode at high ecliptic latitude regions can be improved due to the high redundancy. \\
{\em $^b$} 5$\sigma$ sensitivity without source confusion and sky confusion. In
the case of the \textit{Herschel} \& \textit{SPICA} mission, we commonly use
the 5$\sigma$ sensitivity of \textit{Herschel} mission~\cite{poglit02}. \\
{\em $^c$} Our actual estimations in the LW band are for those in 160 $\mu$m.  We
considered the effect of wider beam at 175 $\mu$m of the LW band.
\end{flushleft}
\end{table*}

\begin{figure*}
  \begin{center}
    \epsfxsize = 6.5cm
    \epsffile{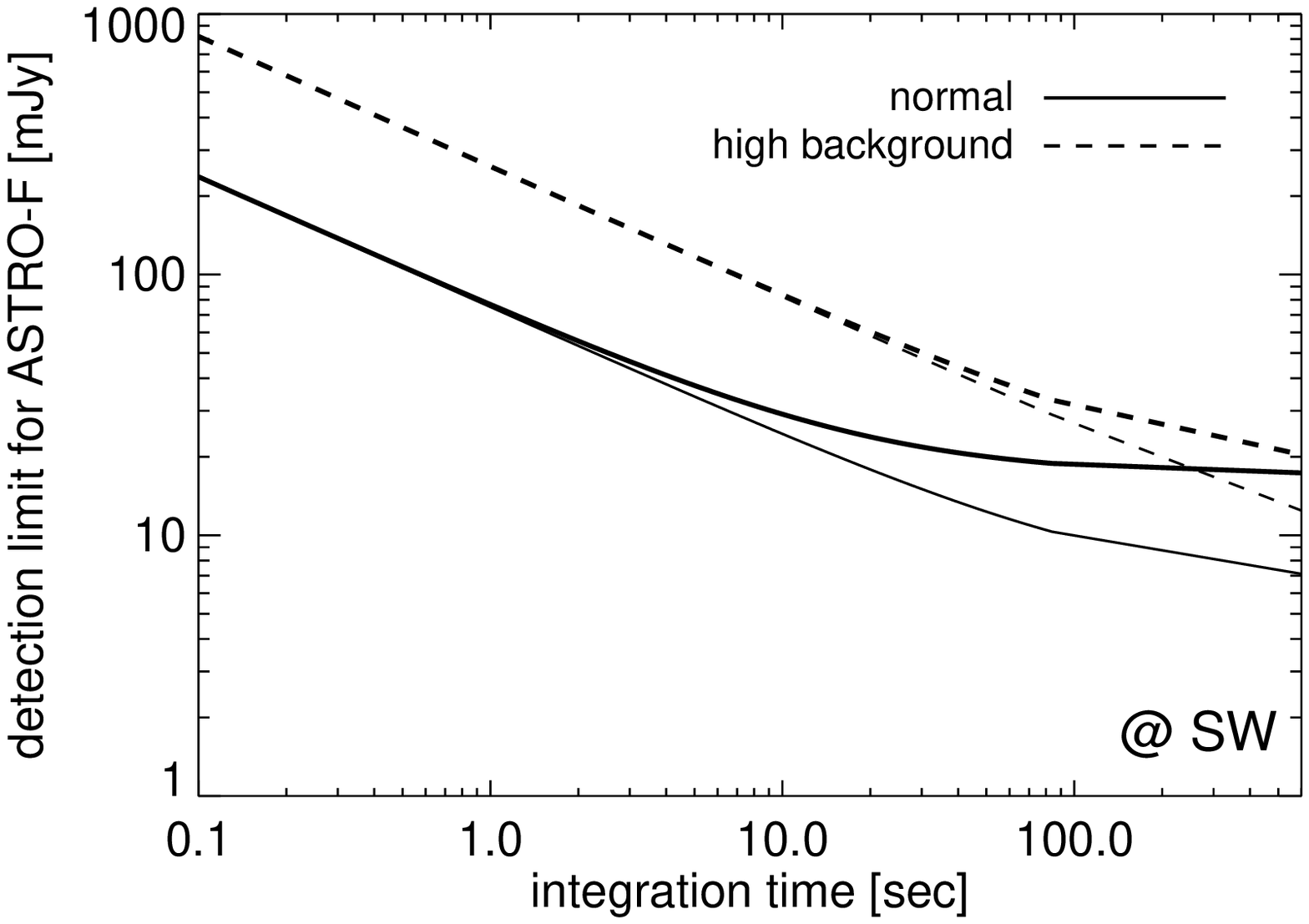}
    \epsfxsize = 6.5cm
    \epsffile{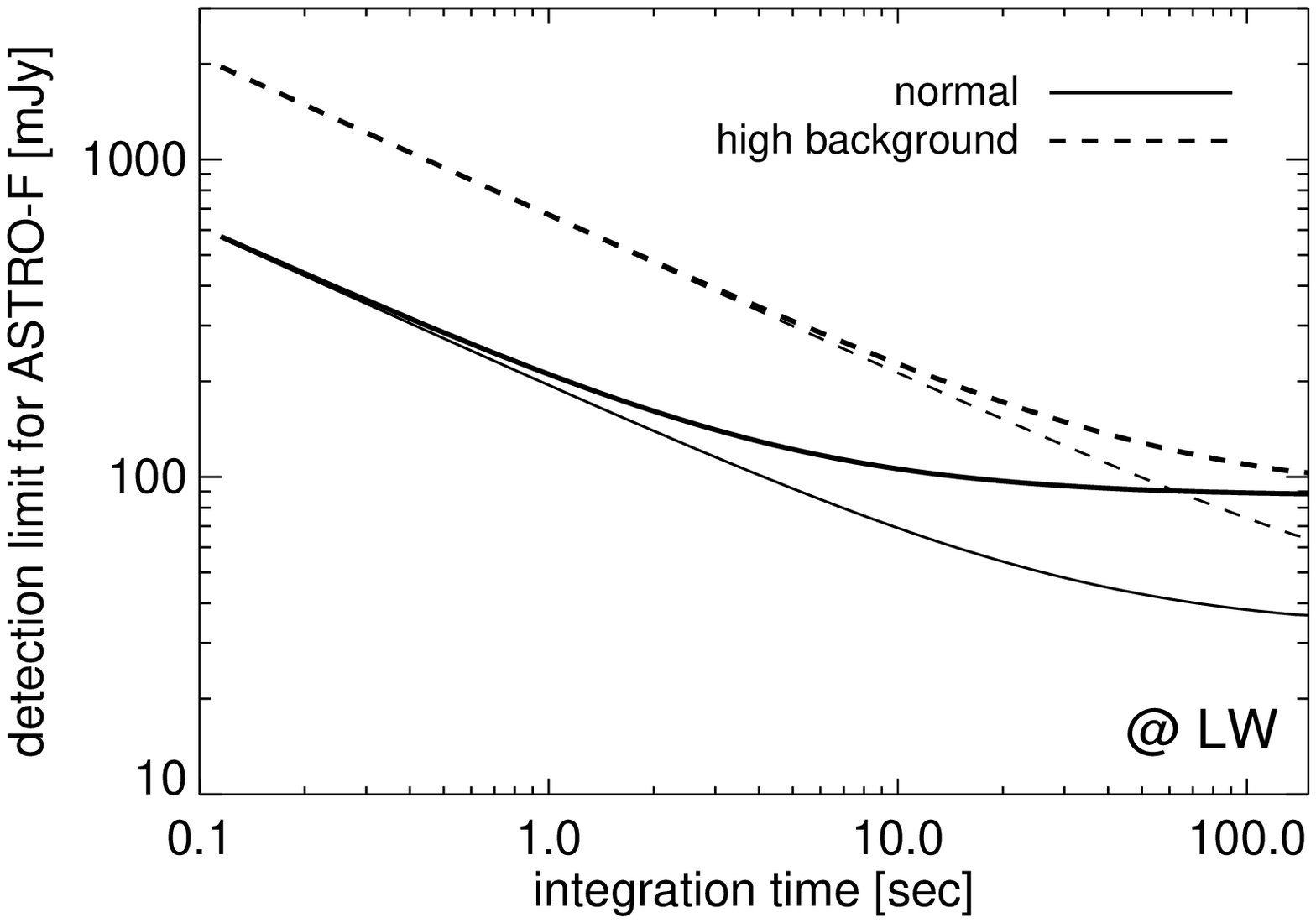}
    \end{center}
    \vspace{-10pt}
   \centering{(a) Detection limits for \textit{AKARI} mission}
  \begin{center}
    \epsfxsize = 6.5cm
    \epsffile{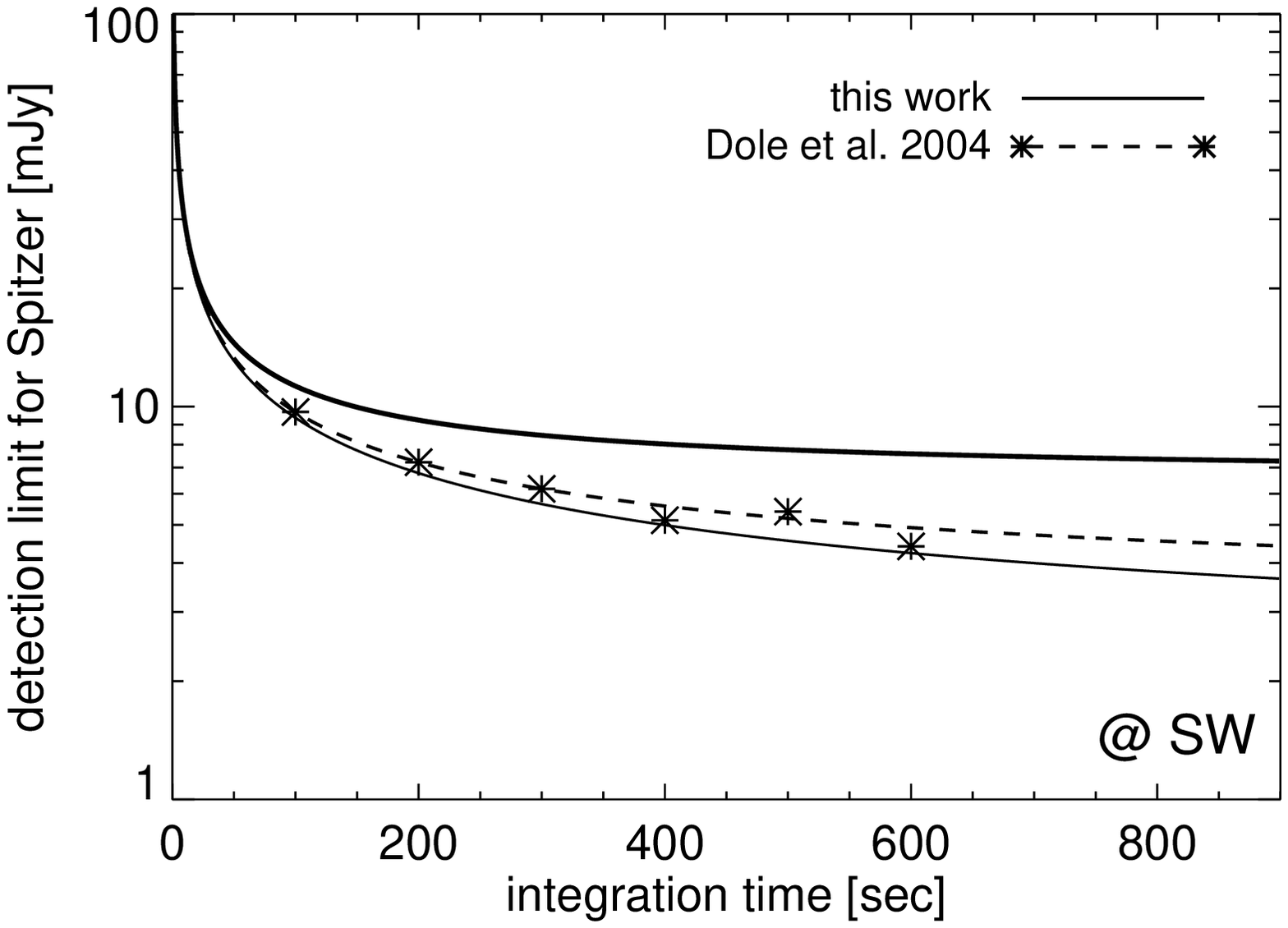}
    \epsfxsize = 6.5cm
    \epsffile{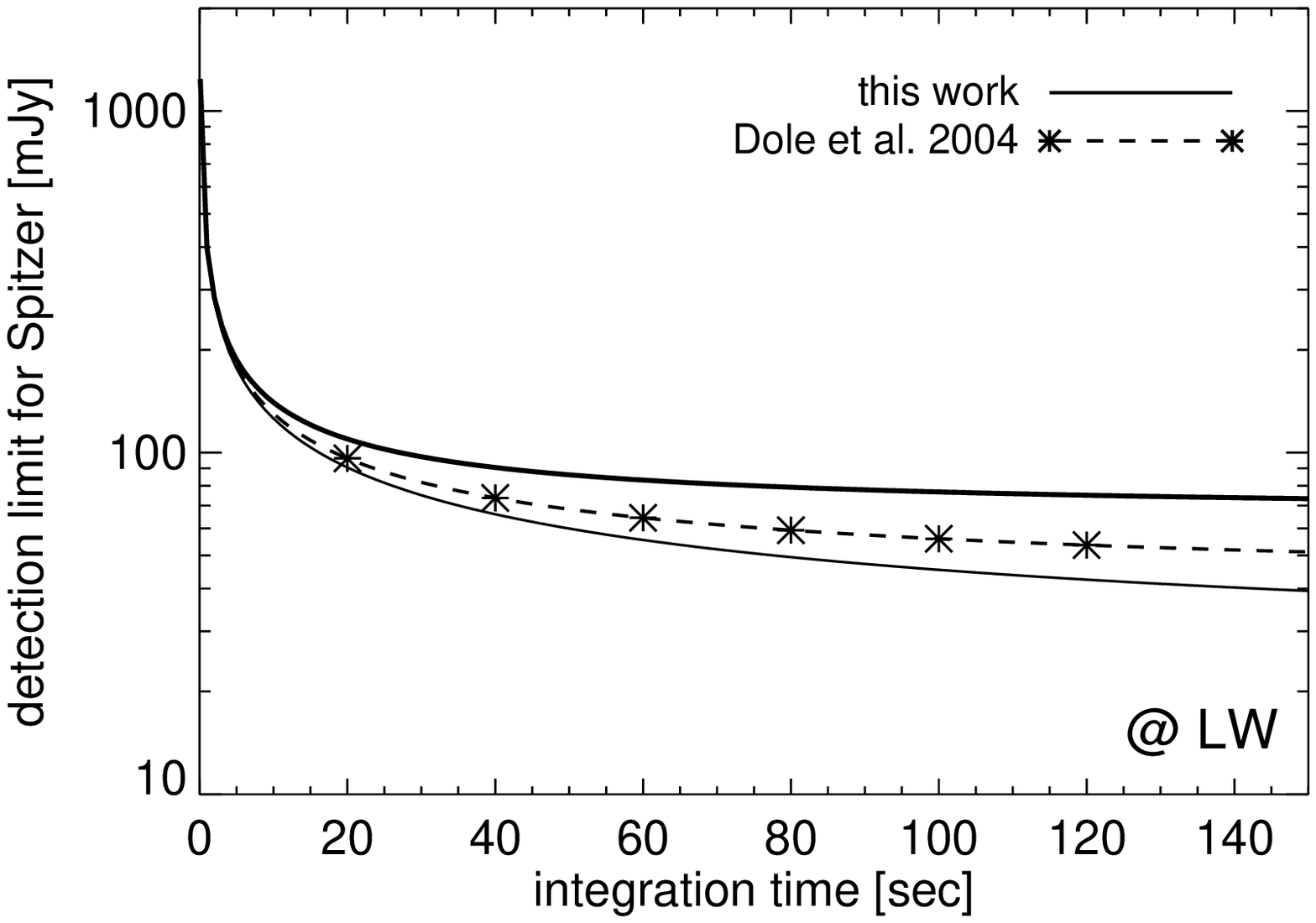}
    \end{center}
    \vspace{-10pt}
   \centering{(b) Detection limits for \textit{Spitzer} mission}
   \caption{Comparison of detection limits with integration time for
   (a) \textit{AKARI} mission and (b) \textit{Spitzer} mission.
   The left panels are for SW band, the right panels for LW band.
   In the case of \textit{AKARI} mission, the solid line shows the detection
   limits over an integration time for normal sampling mode and the dashed
   line shows those for sampling mode in high background region to avoid
   saturation. We plot two evolutionary models. Upper thick line is for the burst
   evolution model and lower thin line for the luminosity evolution model.
   Note that the integration time for single scan is 0.14s and 0.23s for the SW band
   and the LW band, respectively for \textit{AKARI} and the integration time is
   plotted in logarithmic and linear scales for the \textit{AKARI} and
   \textit{Spitzer} missions, respectively.}
   \label{fig_noise_intg}
\end{figure*}

The detection limits as a function of integration time can be an important
indicator of the confusion level. Based upon recent hardware measurements for
the \textit{AKARI} mission and the \textit{Spitzer} results from Dole et al.
(2004b), we compare the detection limits as a function of an integration time
in Figure \ref{fig_noise_intg}. In case of the \textit{AKARI} mission, we also
plot the results of different sampling modes for the high background regions.
The decreasing component of the noise is the usual behavior of noise with
increasing integration time while the flattening component is mainly due to the
confusion. Since  a low cirrus region was assumed in this estimation, the
flattening of the noise results from the source confusion. In the SW band, the
detection limits are mostly determined by the instrumental noise and the source
confusion noise irrespective of the amount of cirrus background except for the
very highest cirrus regions, i.e. near the Galactic plane. However, in the LW
band, we find that the sky confusion affects the noise level severely in the
regions brighter than $\sim$ 20 MJy~sr$^{-1}$ for the \textit{AKARI} mission
and $\sim$ 30 MJy~sr$^{-1}$ for the \textit{Spitzer} mission, respectively (see
also the plot of confusion limits in Fig. \ref{fig_conf_all}). If the different
sampling mode for high background regions available for the \textit{AKARI}
mission is used, the detection limits are mostly limited by instrumental and
photon noises in the SW band unless the exposure time is very long. However, in
the LW band, we found that the sky confusion limits compete with the
instrumental and photon noises even for short exposure observations.

\section{EXPECTED RESULTS}\label{sec:expected}

Based upon two evolutionary (and a no evolution) scenarios of source distributions, we have
estimated the expected confusion limits for various IR missions in Section
\ref{sec:conf_limits}. The two evolutionary models, the luminosity evolution model
and burst evolution model, provide upper and lower limits for the final detection
limits. Here, we discuss the results expected from the two evolutionary models
from our estimated sensitivities.

\subsection{Expected Optimal Redshift Distribution}

\begin{figure*}
  \begin{center}
    \epsfxsize = 8cm
    \epsffile{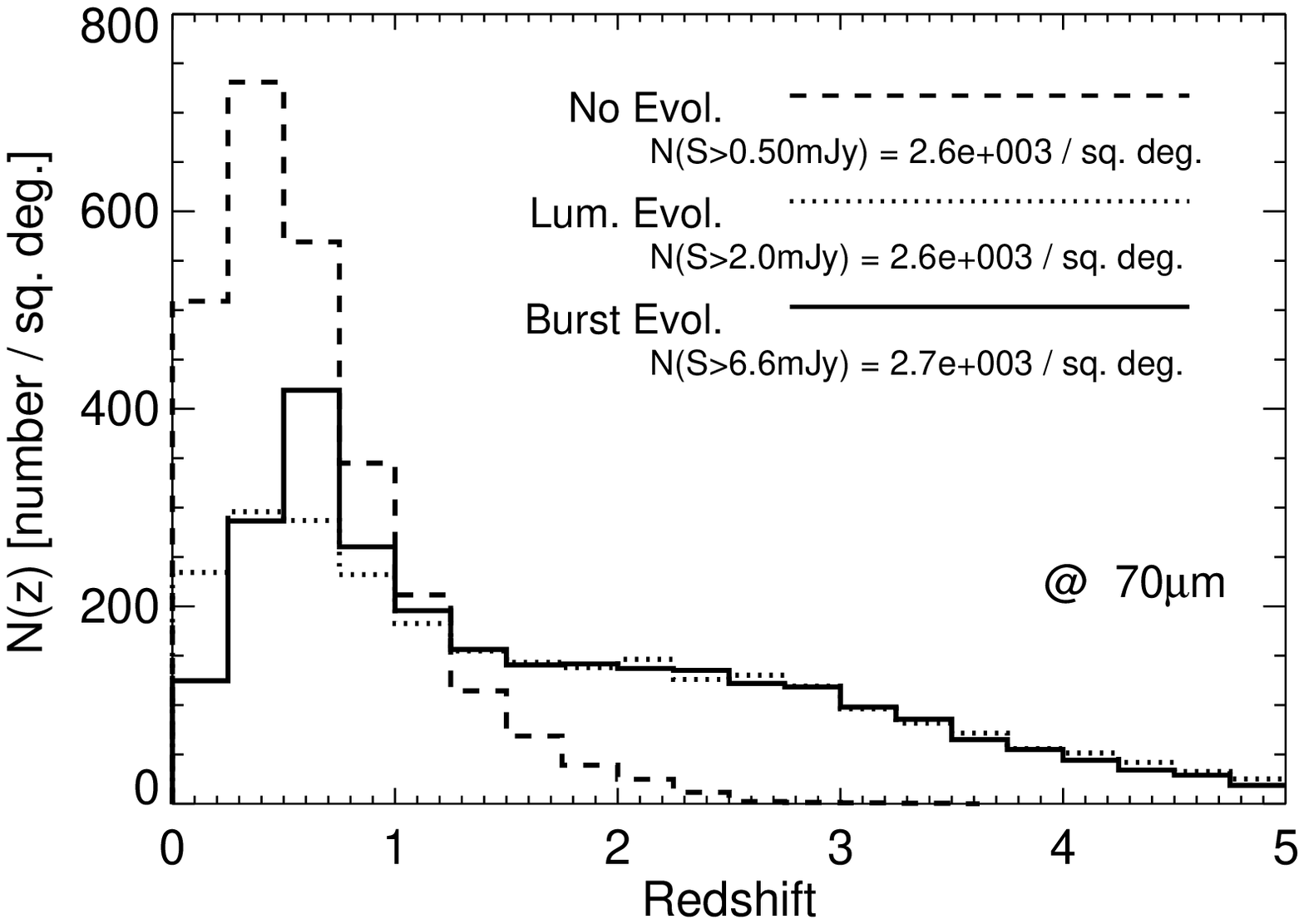}
    \epsfxsize = 8cm
    \epsffile{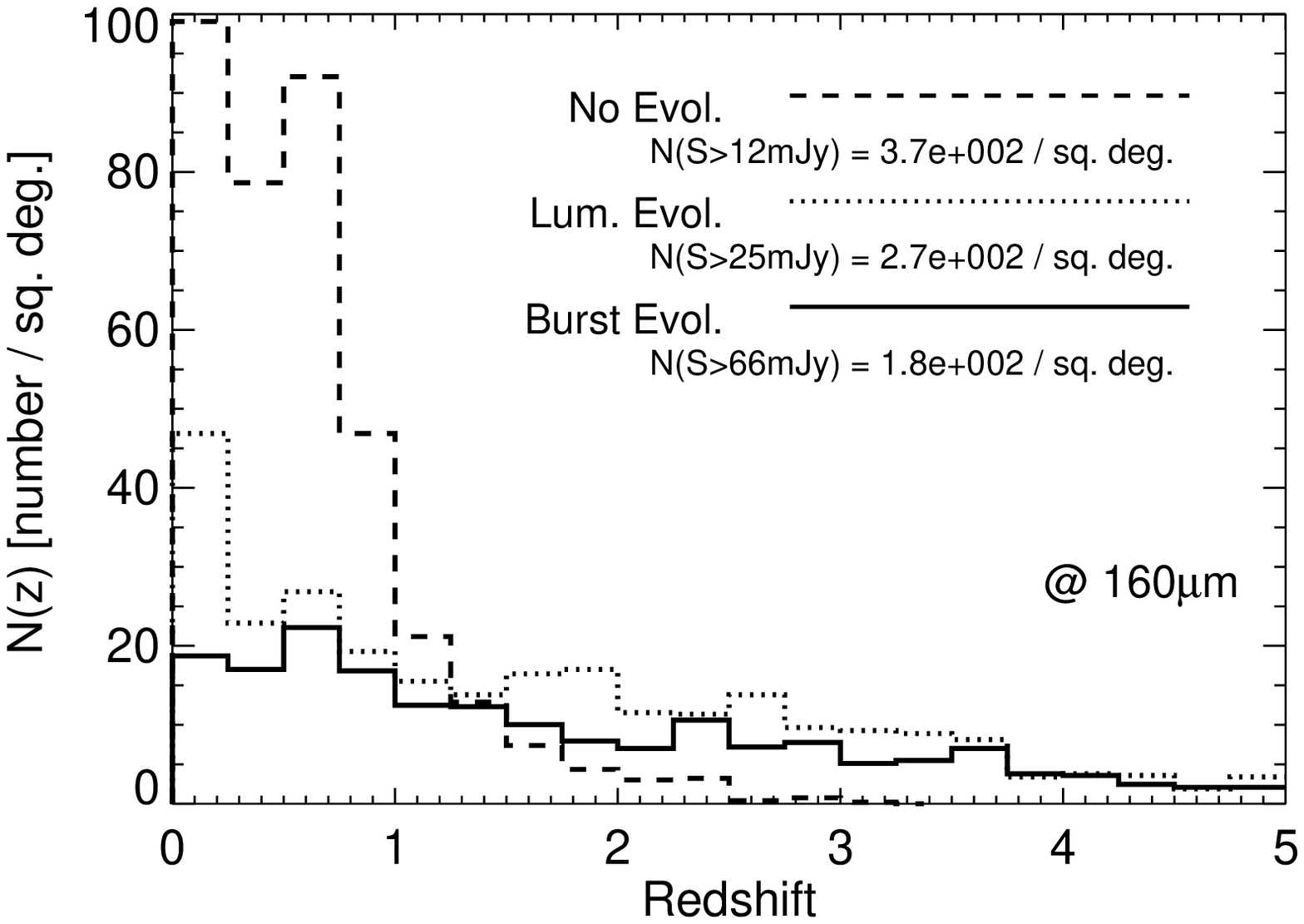}
   \end{center}
    \vspace{-10pt}
   \centering{(a) Redshift distribution for \textit{Spitzer} mission}
  \begin{center}
    \epsfxsize = 8cm
    \epsffile{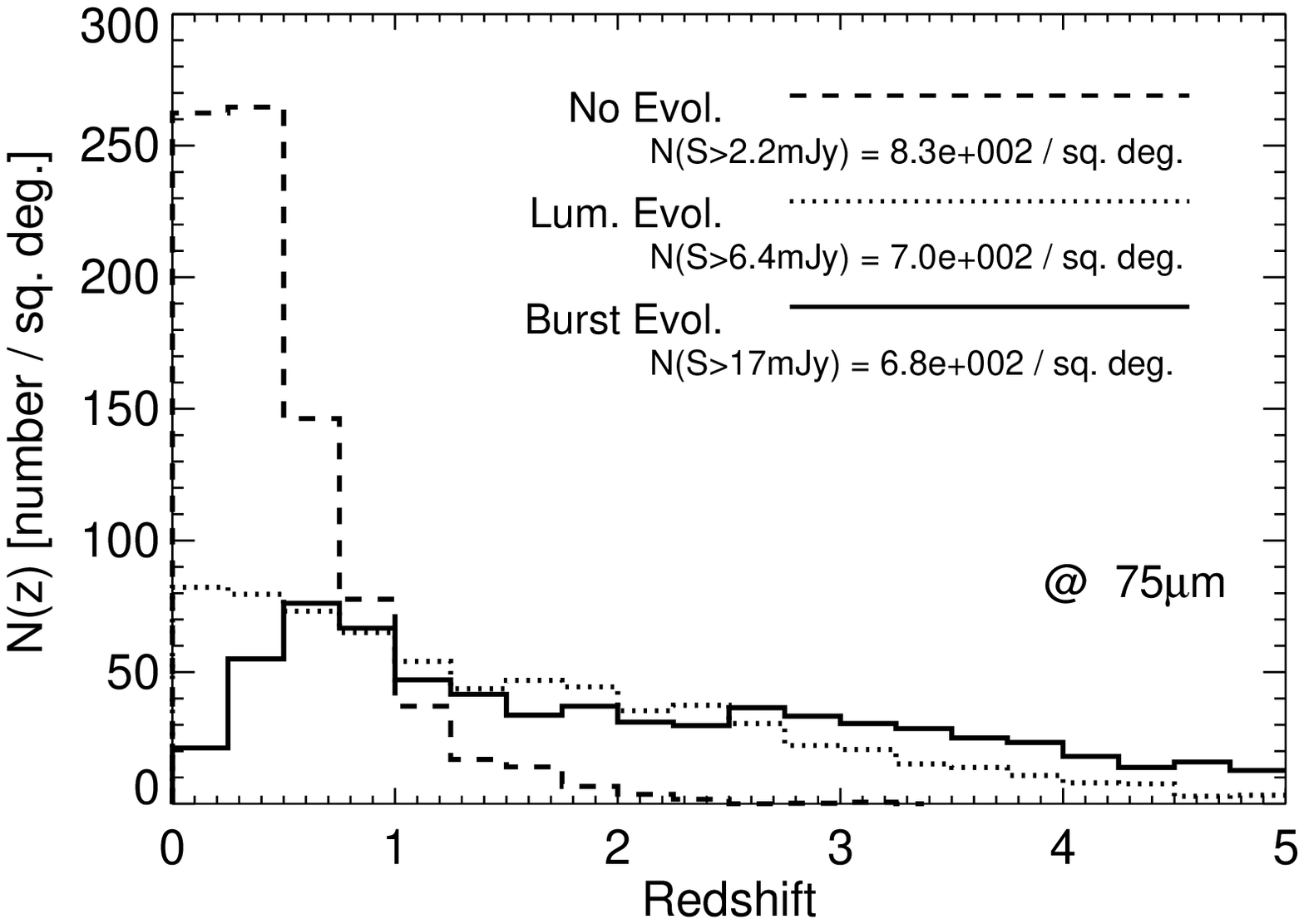}
    \epsfxsize = 8cm
    \epsffile{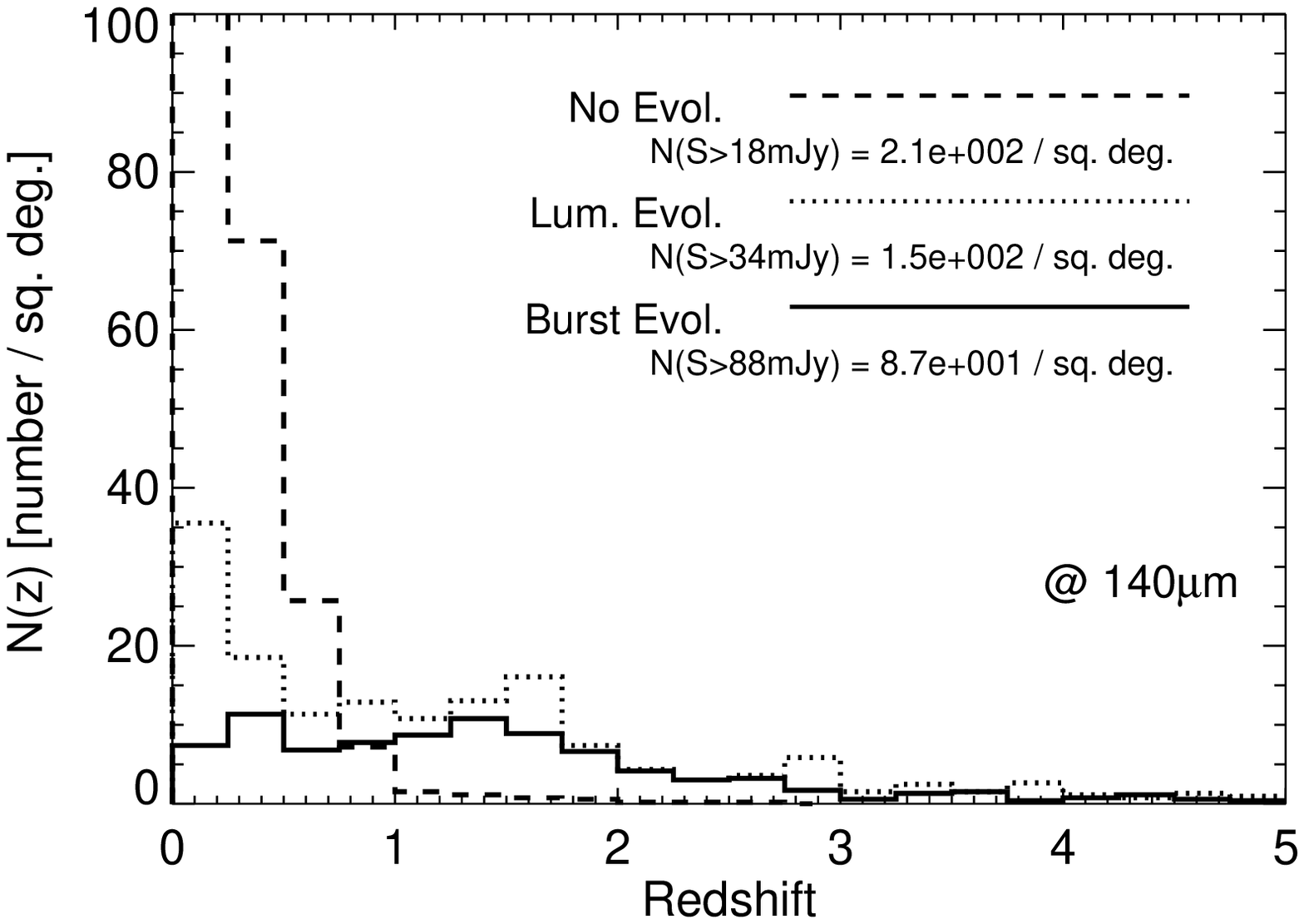}
   \end{center}
    \vspace{-10pt}
   \centering{(b) Redshift distribution for \textit{AKARI} mission}
  \begin{center}
    \epsfxsize = 8cm
    \epsffile{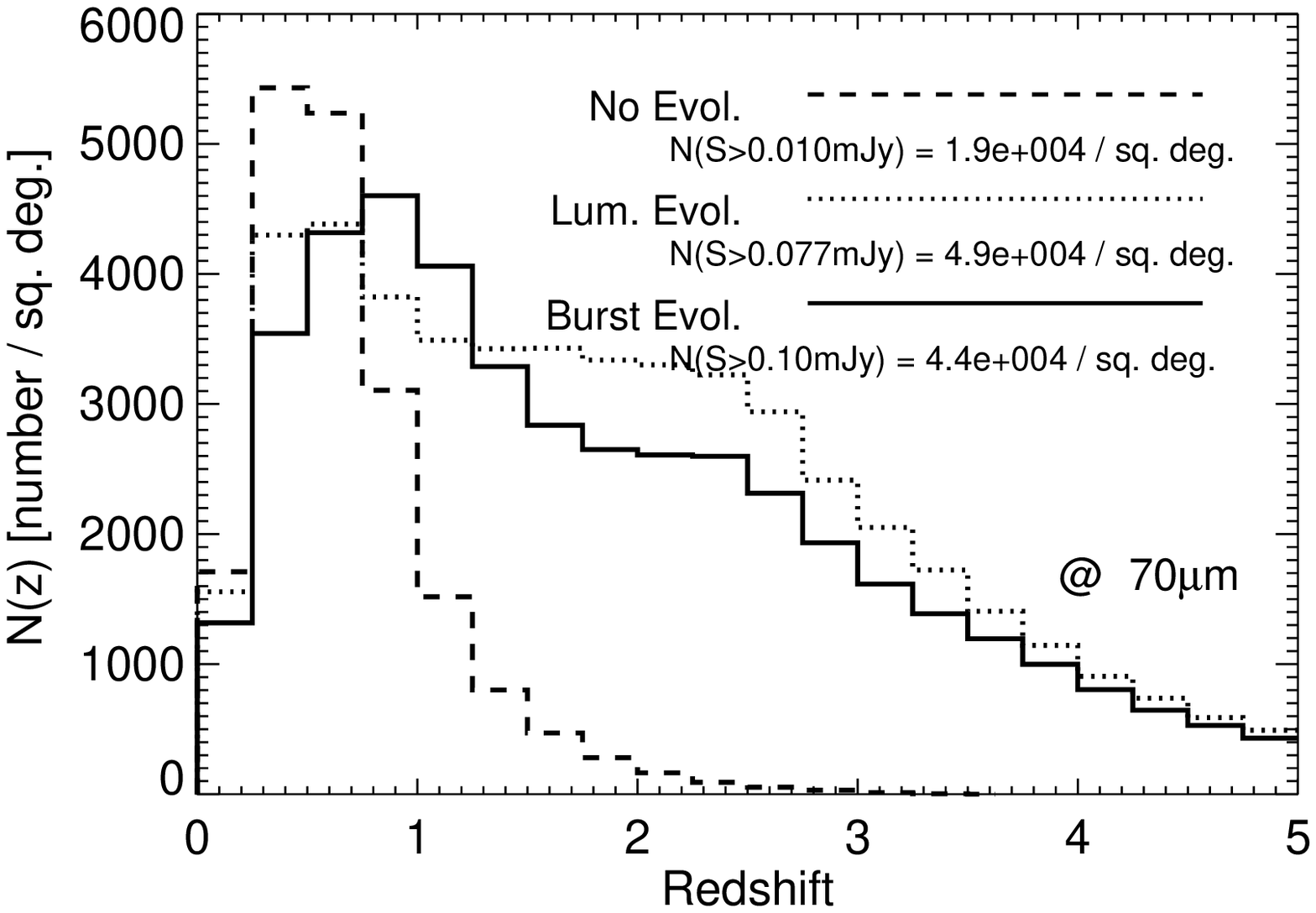}
    \epsfxsize = 8cm
    \epsffile{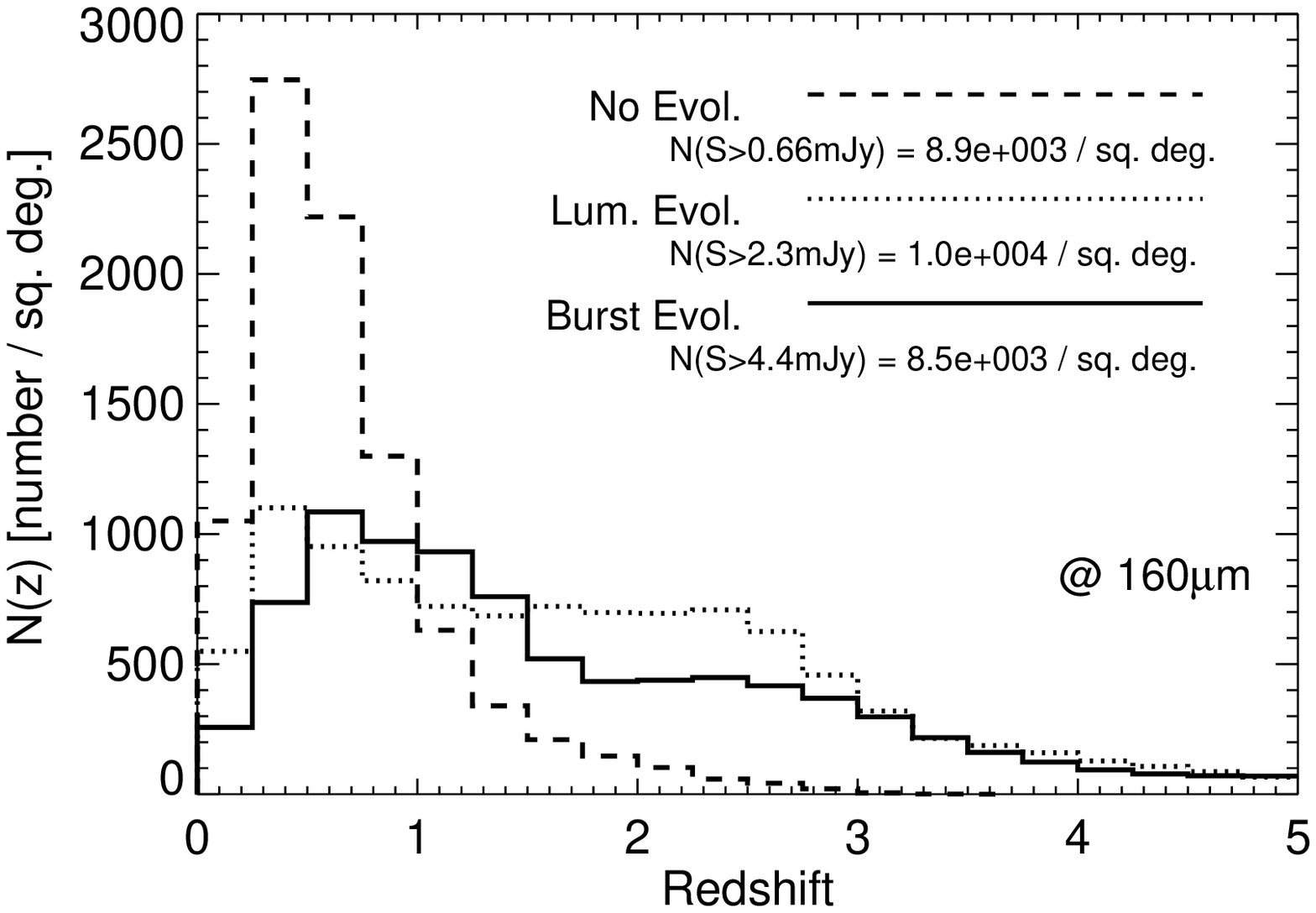}
    \end{center}
    \vspace{-10pt}
   \centering{(c) Redshift distribution for \textit{Herschel} \& \textit{SPICA} mission}
   \caption{Expected optimal redshift distributions for the (a) \textit{Spitzer}, (b) \textit{AKARI}
    and (c) \textit{Herschel} \& \textit{SPICA} missions. The 3 evolutionary
   models are shown for the SW and LW bands respectively at the confusion limit.
   This are the optimal distributions for each mission at the absolute sensitivity
   defined by our analysis.}
   \label{fig_rsdist}
\end{figure*}

Once the realistic detection limits have been determined, we can obtain the
expected optimal confusion limited redshift distribution from the models. The
number-redshift distribution to our limiting flux $S_{\mathrm{lim}}$, can be
extracted directly from our simulated catalogue. In Figure \ref{fig_rsdist}, we
show the expected redshift distributions and expected number of sources per
square degree for each mission at the predicted confusion limit. We consider
this as the optimal redshift distribution for each mission. Since a fairly
large number of sources below $z <$ 1.0 are detected in the LW bands for the no
evolution model, the redshift distribution for each mission has a significant
difference depending on whether there is evolution or not. Though we can find
the peaks in the SW band of the burst evolution model and the LW band of
luminosity evolution model for the \textit{Spitzer} and \textit{AKARI}
missions, we still have difficulty in finding a significant difference between
the two evolutionary scenarios. However, we expect that the large 3.5m aperture
missions (e.g. \textit{Herschel} and \textit{SPICA}) will have advantages over
the 60--90cm aperture missions (\textit{Spitzer} and \textit{AKARI}) in
distinguishing between the evolutionary scenarios due to the larger numbers of
sources detected in the redshift range of 1--3. Since the source confusion
should be more severe for stronger evolutionary models, the total number of
detected sources in the no evolution model is largest in the LW band except for
the \textit{Herschel} and \textit{SPICA} missions.

From Figure \ref{fig_rsdist}, we show that the confusion limit for the
\textit{Spitzer} and \textit{AKARI} missions, restricts the majority of the
sources to redshifts of around or less than unity. For these missions, we could
expect of the order of a few tens of sources per redshift bin per square degree
out to redshifts of 2-3 (Note that the MIPS 70 $\mu$m band is more sensitive in
theory but suffers from detector array problems in practice). However, for the
\textit{Herschel} and \textit{SPICA} missions, we could expect 500-1000s of
sources per square degree per redshift bin out to redshifts of 2-3. This would
allow direct probing of the dusty starburst populations detected at 24 $\mu$m
with \textit{Spitzer} \cite{papovich04,rodighiero05} and at 850$\mu$m with
\textit{SCUBA} [e.g. Chapman et al. \shortcite{chapman03}] resulting in the
construction of much more accurate far-IR luminosity functions out to these
redshifts with large number statistics.

\subsection{Expected Cosmic Far-Infrared Background}

The measurements of the infrared background have been carried out in several
wavelength bands. Also, there has been rapid progress in resolving a
significant fraction of the background with deep galaxy counts at infrared
wavelengths.

The flux from extragalactic sources below the detection limits create
fluctuations in the background. The Cosmic Far-Infrared Background (CFIRB)
intensity $I_{\rm CFIRB}$ produced by all sources with the flux below the limiting
flux $S_{\rm lim}$, is obtained from:
\begin{equation}
  I_{\rm CFIRB} = \int^{S_{\rm lim}}_{0} S~{dN \over dS}~dS.
\label{eqn_int_cfirb}
\end{equation}
In addition, the CFIRB fluctuations $P_{\rm CFIRB}$ from sources with a
random distribution below a given detection limit $S_{\rm lim}$ can be
estimated as follows,
\begin{equation}
  P_{\rm CFIRB} = \int^{S_{\mathrm{lim}}}_{0} S^2~{dN \over dS}~dS.
\label{eqn_fluc_cfirb}
\end{equation}
The detection limits $S_{\mathrm{lim}}$ can be found in Table
\ref{tab_final_sc}.

We list the expected CFIRB intensity and resolved fraction for each
mission in Table \ref{tab_int_cfirb}. Dole et al. \shortcite{dole04b} predicted
that the \textit{Spitzer} mission can resolve 18\% of the CFIRB at 160 $\mu$m
from their source count model. According to our estimation with our source count
model, the resolved CFIRB is expected to be 9 -- 17\% of the total CFIRB. As
shown in Table \ref{tab_int_cfirb}, the \textit{Herschel} \& \textit{SPICA}
missions will resolve a much larger fraction of the CFIRB, i.e. 90 -- 94\% in
the SW band and 60 -- 72\% in the LW band, compared to other missions.

\begin{table*}
\centering \caption{Expected intensity, fluctuation, and resolved fraction of
CFIRB for each mission. The upper flux is set to be the final confusion limit.}
\label{tab_int_cfirb} \vspace{5pt}
\begin{tabular}{@{}ccccccccc}
\hline\vspace{-5pt} \\
& \multicolumn{4}{c}{Luminosity evolution} & \multicolumn{4}{c}{Burst evolution} \vspace{5pt} \\
 & \multicolumn{2}{c}{SW} & \multicolumn{2}{c}{LW} & \multicolumn{2}{c}{SW} & \multicolumn{2}{c}{LW}
 \vspace{5pt}\\
  &  Intensity & Resolved & Intensity & Resolved & Intensity & Resolved &
Intensity & Resolved \vspace{5pt} \\
 Space Mission & (MJy/sr) & (\%) & (MJy/sr) &
(\%) & (MJy/sr) & (\%) & {(MJy/sr)} & (\%) \vspace{5pt}
\\\hline \vspace{-10pt}
\\ \textit{Spitzer} & 0.063 & 48 & 0.33 & 17 & 0.12 & 52 & 0.69 & 9 \vspace{5pt}
\\ \textit{AKARI} & 0.094 & 32 & 0.31 & 14 & 0.21 & 27 & 0.66 & 7 \vspace{5pt}
\\ \textit{Herschel} \& \textit{SPICA} & 0.013 & 90 & 0.16 & 60 & 0.015 & 94 & 0.21 & 72 \vspace{5pt}
\\ \hline
\end{tabular}
\end{table*}

\begin{figure}
  \begin{center}
    \epsfxsize = 8cm
    \epsffile{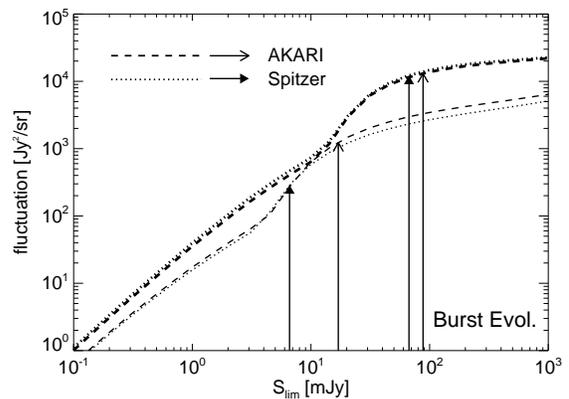}
    \end{center}
   \caption{Expected CFIRB fluctuations for the burst evolution model. Thick
   lines are for the LW band and thin lines for the SW band. Two arrows show the
   expected CFIRB fluctuation corresponding to the detection limits.}
   \label{fig_fluc_ulig}
\end{figure}

Table \ref{tab_fluc_cfirb} lists the expected CFIRB fluctuations for each
mission. We can not find a significant difference in the fluctuation levels of
the \textit{Spitzer} and the \textit{AKARI} missions in the LW band since the
detection limits already approach a flux range with a monotonic increase of
fluctuation (see Fig. \ref{fig_fluc_ulig} for the burst evolution model). In
the far-IR range, Lagache \& Puget \shortcite{lagache00b} and Matsuhara et al.
\shortcite{matsuhara00} have studied the detection of the CFIRB fluctuation
with \textit{ISO} data in Marano 1 region and the Lockman Hole, respectively.
For comparison, we list the fluctuation estimated from our model for the
\textit{ISO} mission in Table \ref{tab_fluc_comp}. We find that the observed
fluctuations are mostly located between the results from our two evolution models
except for the observations at 90 $\mu$m. Comparing with the model of Lagache,
Dole \& Puget \shortcite{lagache03}, our estimated fluctuations from the burst
evolution model are in good agreement with those from their model.

\begin {table}
\centering \caption{Expected CFIRB fluctuations for each mission.}
\label{tab_fluc_cfirb} \vspace{5pt}
\begin{tabular}{@{}crlrl}
\hline\vspace{-5pt} \\
& \multicolumn{2}{c}{Luminosity evolution} & \multicolumn{2}{c}{Burst evolution} \vspace{5pt} \\
 & \multicolumn{2}{c}{(Jy$^2$/sr)} & \multicolumn{2}{c}{(Jy$^2$/sr)} \vspace{5pt} \\
Space Mission & ~~~~~~~~SW & LW & ~~~~SW & LW \vspace{5pt}
\\\hline \vspace{-10pt}
\\ \textit{Spitzer} & 35 & 1600 & 290 & 12000 \vspace{5pt}
\\ \textit{AKARI} & 130 & 1800 & 1200 & 13000 \vspace{5pt}
\\ \textit{Herschel} \& \textit{SPICA} & 0.4 & 120 & 0.6 & 290 \vspace{5pt}
\\ \hline
\end{tabular}
\end{table}

\begin {table*}
\centering \caption{Comparison of CFIRB fluctuations for \textit{ISO} mission.}
\label{tab_fluc_comp} \vspace{5pt}
\begin{tabular}{@{}cccccc}
\hline\vspace{-5pt} \\
$\lambda$ & $\theta$ & $S_{\rm max}$ & Observations & Predicted $^a$ & Predicted (this work) $^b$ \vspace{5pt} \\
 ($\mu$m) & (arcmin) & (mJy) & (Jy$^2$/sr) & (Jy$^2$/sr) & (Jy$^2$/sr) \vspace{5pt}
\\\hline \vspace{-10pt}
\\ 90 & 0.4 $-$ 20 & 150 & 13000 $\pm$ 3000~$^c$ & 5300 & 2100 $-$ 7200 \vspace{5pt}
\\ 170 & 0.6 $-$ 4 & 100 & 7400~$^d$ & 12000 & 3800 $-$ 14000 \vspace{5pt}
\\ 170 & 0.6 $-$ 20 & 250 & 12000 $\pm$ 2000~$^c$ & 16000 & 5500 $-$ 18000 \vspace{5pt}
\\ \hline
\end{tabular}
\medskip
\begin{flushleft}
{\em $^a$} Model from Lagache, Dole \& Puget \shortcite{lagache03}.\\
{\em $^b$} Lower limit is estimated from luminosity evolution model and upper
limit from burst evolution model. \\
{\em $^c$} Observational analysis from Matsuhara et al. \shortcite{matsuhara00}.\\
{\em $^d$} Observational analysis from Lagache \& Puget \shortcite{lagache00b}.
\end{flushleft}
\end{table*}

\begin{figure}
  \begin{center}
    \epsfxsize = 8cm
    \epsfysize = 6.0cm
    \epsffile{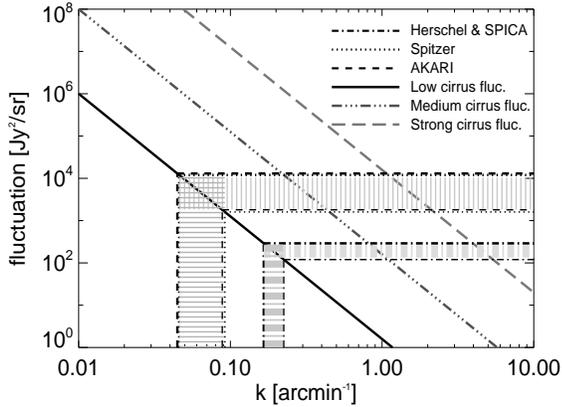}
    \end{center}
   \caption{Comparison between expected CFIRB and cirrus fluctuation for each
   mission in the LW band. The solid line shows the power spectrum of cirrus
   emission at high Galactic latitude. For comparison, we also plot the power
   spectrum in medium and high cirrus regions. The shaded area shows fluctuations
   and corresponding spatial frequencies covering two evolution models at high
   Galactic latitude for each mission. The lower limit is for luminosity evolution
   and the upper limit for burst evolution.}
   \label{fig_cfirb_fluc}
\end{figure}

\begin{figure}
  \begin{center}
    \epsfxsize = 8cm
    \epsfysize = 6.0cm
    \epsffile{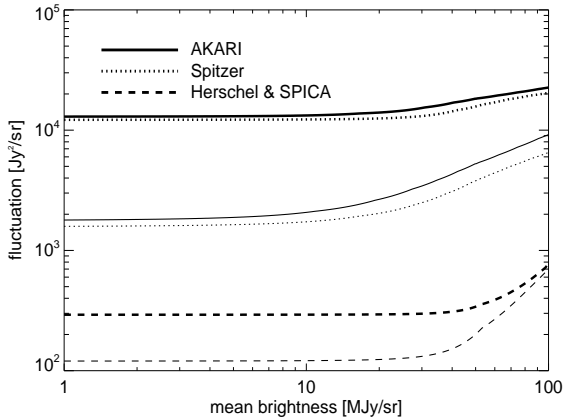}
    \end{center}
   \caption{Expected CFIRB fluctuation as a function of cirrus mean brightness
   in the LW band. The two lines plotted for each mission are for the burst evolution model
   (upper line) and for the luminosity evolution model (lower line).}
   \label{fig_cfirb_cirrus}
\end{figure}

The power spectrum of cirrus emission at high Galactic latitudes ($>$ 80
degree) also has a fluctuation of about 10$^6$ Jy$^2$/sr at 0.01 arcmin$^{-1}$
at 160 $\mu$m with a power index of -2.9 $\pm$ 0.5. In order to distinguish
the CFIRB fluctuations from the observed data that include both CFIRB and
cirrus emission effectively, we require an area larger than 115, 125, and 19
arcmin$^2$ for the \textit{Spitzer}, \textit{AKARI}, and the \textit{Herschel} \&
\textit{SPICA} missions, respectively for the case of luminosity evolution
model. If we adopt the burst evolution model, the minimum required area for the
resolution of the estimated power spectrum should increase to 463, 490, 36
arcmin$^2$ for the \textit{Spitzer}, \textit{AKARI}, and the
\textit{Herschel} \& \textit{SPICA} missions, respectively. The shaded area in
Figure \ref{fig_cfirb_fluc} covers the fluctuations for both evolutionary
models. In this estimation, we do not consider the clustering of extragalactic
sources. As sky confusion noise increases in the low Galactic latitude regions,
detection limits should also degrade, and the CFIRB fluctuation becomes larger
(see Figure \ref{fig_cfirb_cirrus}). As can be seen in Figure
\ref{fig_cfirb_cirrus}, the fluctuations show a monotonic increase to medium
cirrus regions. Since the fluctuation power in low Galactic latitude regions
should be more than 10$^{10}$ Jy$^2$/sr at 0.01 arcmin$^{-1}$ at 160
$\mu$m, it is difficult to extract the CFIRB fluctuations from the analysis of
power spectrum [Note that this CFIRB fluctuation is expected to have spatial
structure below 1 arcmin$^{-1}$ (see Figure \ref{fig_cfirb_fluc})]. However, we
expect to detect CFIRB fluctuations in most of the low-to-medium cirrus
regions.

\section{SUMMARY AND CONCLUSIONS}\label{sec:conclusions}

In order to probe the confusion limit for infrared observations, we generated
source catalogues assuming a concordance (i.e. flat, dark energy dominated)
cosmological world model ($H_{o}=72$, $\Omega=0.3$, $\Lambda=0.7$) for 2
evolutionary scenarios defined as the luminosity evolution model and burst
model following Pearson \shortcite{cpp96} and Pearson et al. \shortcite{cpp01},
respectively. We also considered the sky confusion due to the infrared cirrus.
Though the sky confusion is not a dominant noise source at high Galactic
latitudes, we should take into account the effects of fluctuations of the sky
brightness for large area surveys.

Based upon the fluctuation analysis and photometry on simulated images, we find
a composite estimator that represents the source confusion well. From our
analysis of source confusion and source distribution models including galaxy
evolution, we have estimated final confusion limits of 2.0 -- 6.6 mJy and 25 --
67 mJy at 70 $\mu$m and 160 $\mu$m for the \textit{Spitzer} mission, 6.4 -- 17
mJy and 34 -- 88 mJy at 75 $\mu$m and 140 $\mu$m for the \textit{AKARI}
mission, 0.077 -- 0.10 mJy and 2.3 -- 4.4 mJy at 70 $\mu$m and 160 $\mu$m for
the \textit{SPICA} mission, and 0.077 -- 0.10 mJy and 2.6 -- 5.1 mJy at 70
$\mu$m and 175 $\mu$m for the \textit{Herschel} mission in low cirrus regions.
If the source distribution follows the evolutionary models, the current and
planned infrared missions will be mostly limited by source confusion. Other
components affecting confusion are the fluctuation from the zodiacal light,
asteroids, and clustering of sources. As yet the fluctuation of the zodiacal
light on small scales has not been observed although we may expect this
information from the current or next generation of space missions. Based upon
the number distribution \cite{tedesco05} and SED \cite{muller98,kim03} of
asteroids, we roughly estimated the effect of source confusion from asteroids.
Since the slope of the flux-number distribution is shallower than that of
Euclidean space and the number density at any given flux is much smaller than
that of our evolutionary source distributions, we expect that the effect from
asteroids near the ecliptic plane is not severe, i.e., less than 2\% and 0.5\%
in the SW and LW bands, respectively. The fluctuation from the clustering of
sources results in a 10 $\sim$ 15\% increase in the confusion
\cite{take04,negre04}, and can be comparable to the sky confusion in regions of
medium cirrus brightness in the LW band. This effect of clustering on the
confusion should be detectable in the SW band or in regions of low cirrus
brightness. Using our study for sky confusion and source confusion, we have
generated all sky maps for final confusion limits. These maps are shown in
Appendix \ref{app:conf_ass}.

We have also obtained the optimal (confusion limited) redshift distribution
from each source count model. The redshift distributions for each mission show
significant differences between the no evolution scenario and evolutionary
models. However, in order to distinguish between the two evolution scenarios,
higher sensitivity/resolution missions (e.g., \textit{Herschel} and
\textit{SPICA}) are required compared to the present, relatively low resolution
missions. We have estimated that the CFIRB will be resolved to 48 -- 52\% and 9
-- 17\% at 70 $\mu$m and 160 $\mu$m for the \textit{Spitzer} mission, 27 --
32\% and 7 -- 14\% at 75 $\mu$m and 140 $\mu$m for the \textit{AKARI} mission,
and 90 -- 94\% and 60 -- 72\% at 70 $\mu$m and 160 $\mu$m for the
\textit{Herschel} \& \textit{SPICA} missions. We also found that we can detect
the CFIRB fluctuations in most low-to-medium cirrus regions.

\section*{Acknowledgment}
We thank an anonymous referee for valuable comments. This work was financially
supported in part by the KOSEF Grant R14-2002-058-01000-0. CPP acknowledges
support from the Japan Society for the Promotion of Science.

\bigskip



\begin{appendix}

\section{PHOTOMETRIC RESULTS FOR SIMULATED DATA} \label{app:phot_results}

We have carried out photometry on our simulated images under various
conditions. The results of the photometry is shown below.

Fig. \ref{fig_phot_all_nobgr} shows the photometric results for each mission
considering only source confusion. The results for the no evolution model are
clearly segregated from those of the evolution models due to the low source
confusion. We also found that the completeness and the reliability fall rapidly
since the source confusion is more severe for relatively low resolution
observations such as is the case for the LW bands of
\textit{AKARI} and \textit{Spitzer}.

\begin{figure*}
  \hspace{3pt}
  \begin{center}
    \epsfxsize = 13cm
    \epsffile{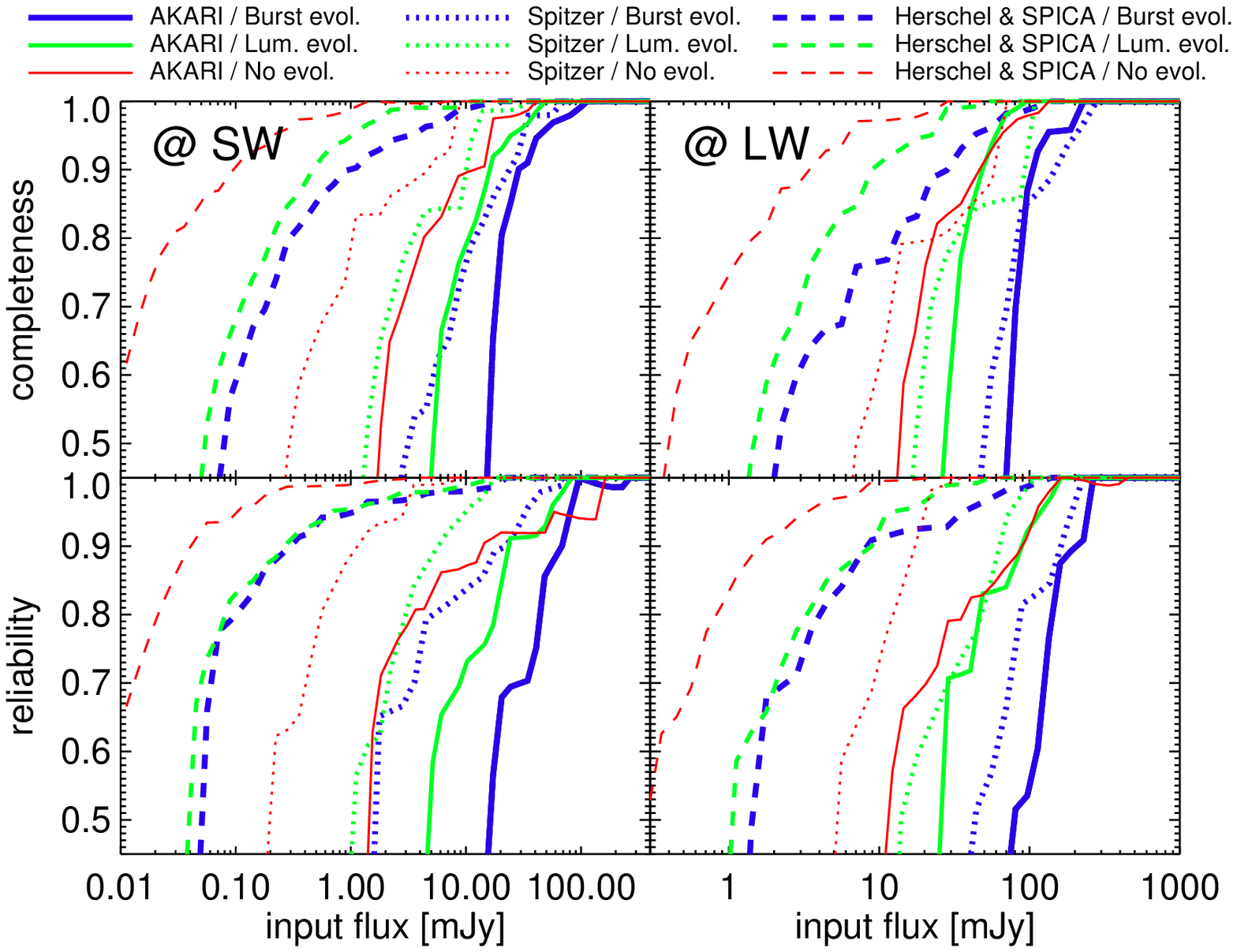}
    \end{center}
   \caption{Photometric results for each mission considering only source confusion.
   Upper panels show the completeness and lower panels the reliability.
   Left and right panels show the SW band and LW band, respectively. A lighter
   and thinner line for the same line style means the result for a different source
   distribution evolutionary model.}
   \label{fig_phot_all_nobgr}
\end{figure*}

\begin{figure}
  \begin{center}
    \epsfxsize = 8.7cm
    \epsffile{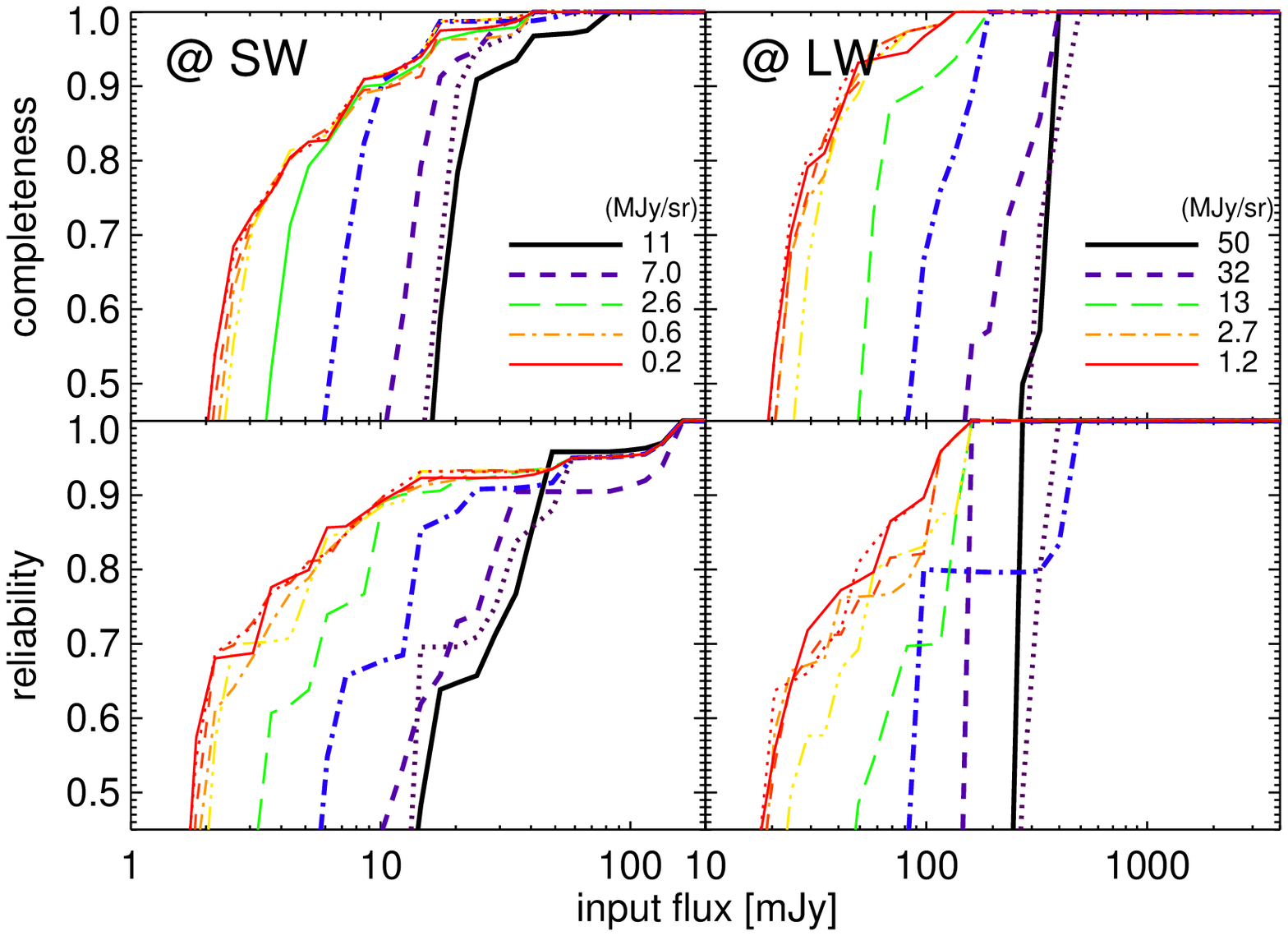}
    \end{center}
   \caption{Photometric results for the \textit{AKARI} mission with the no evolution
   model. Upper panels show the completeness for various cirrus backgrounds and
   lower panels, the reliability. Left and right panels show the results for the
   SW band and LW band, respectively. Each line represents the photometric
   result for each cirrus background. Due to the equivalent or excessive sky confusion
   compared to the source confusion, each result shows a clear difference.}
   \label{fig_phot_fis_noevol}
\end{figure}

Fig. \ref{fig_phot_fis_noevol} shows the photometric results for the
\textit{AKARI} mission with the no evolution model and various levels of
cirrus background. Since the contribution from the sky confusion in the LW band
is comparable to the source confusion (see the top-left panel of Fig.
\ref{fig_conf_all}), the completeness and reliability are well segregated as
the cirrus background increases.

\begin{figure*}
  \begin{center}
    \epsfxsize = 13cm
    \epsffile{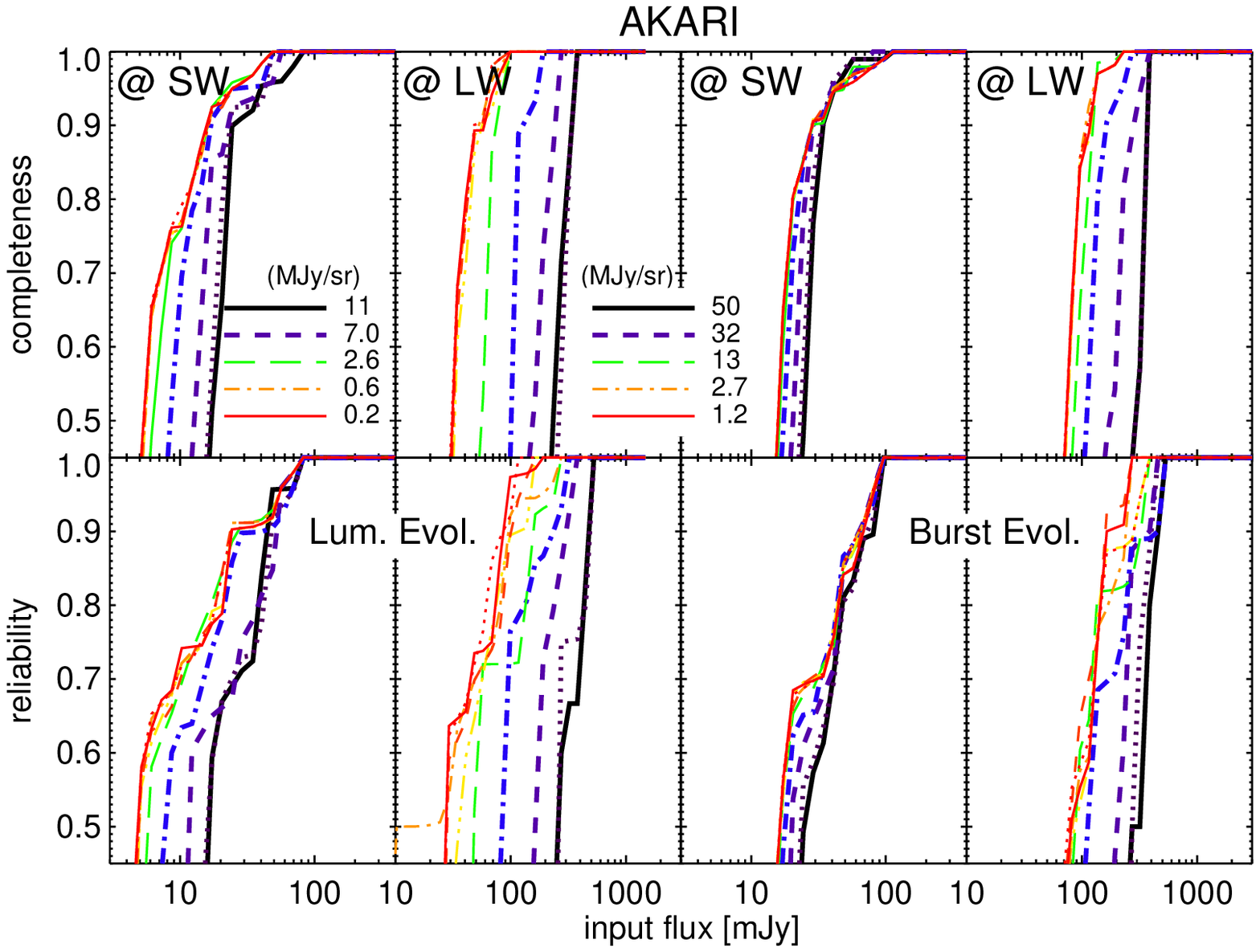}
    \end{center}
   \caption{Photometric results for the \textit{AKARI} mission with the two evolutionary
   models. Upper panels show the completeness for various cirrus backgrounds and
   lower panels, the reliability. Left and right two panels show the results for the
   luminosity evolution and burst evolution models, respectively. Each line represents
   the photometric result for each cirrus background. Due to the equivalent or
   excessive sky confusion compared to the source confusion, each result shows
   a clear difference.}
   \label{fig_phot_fis}
\end{figure*}

Fig. \ref{fig_phot_fis} shows the photometric results for the \textit{AKARI}
mission for the two evolutionary models and various levels of cirrus background.
Since the contribution from the source confusion in the LW band is stronger
than that in no evolution model (Fig. \ref{fig_phot_fis_noevol}, we find
that the plots of completeness and reliability gather closer compared to
those in the no evolution model.

\begin{figure*}
  \begin{center}
    \epsfxsize = 13cm
    \epsffile{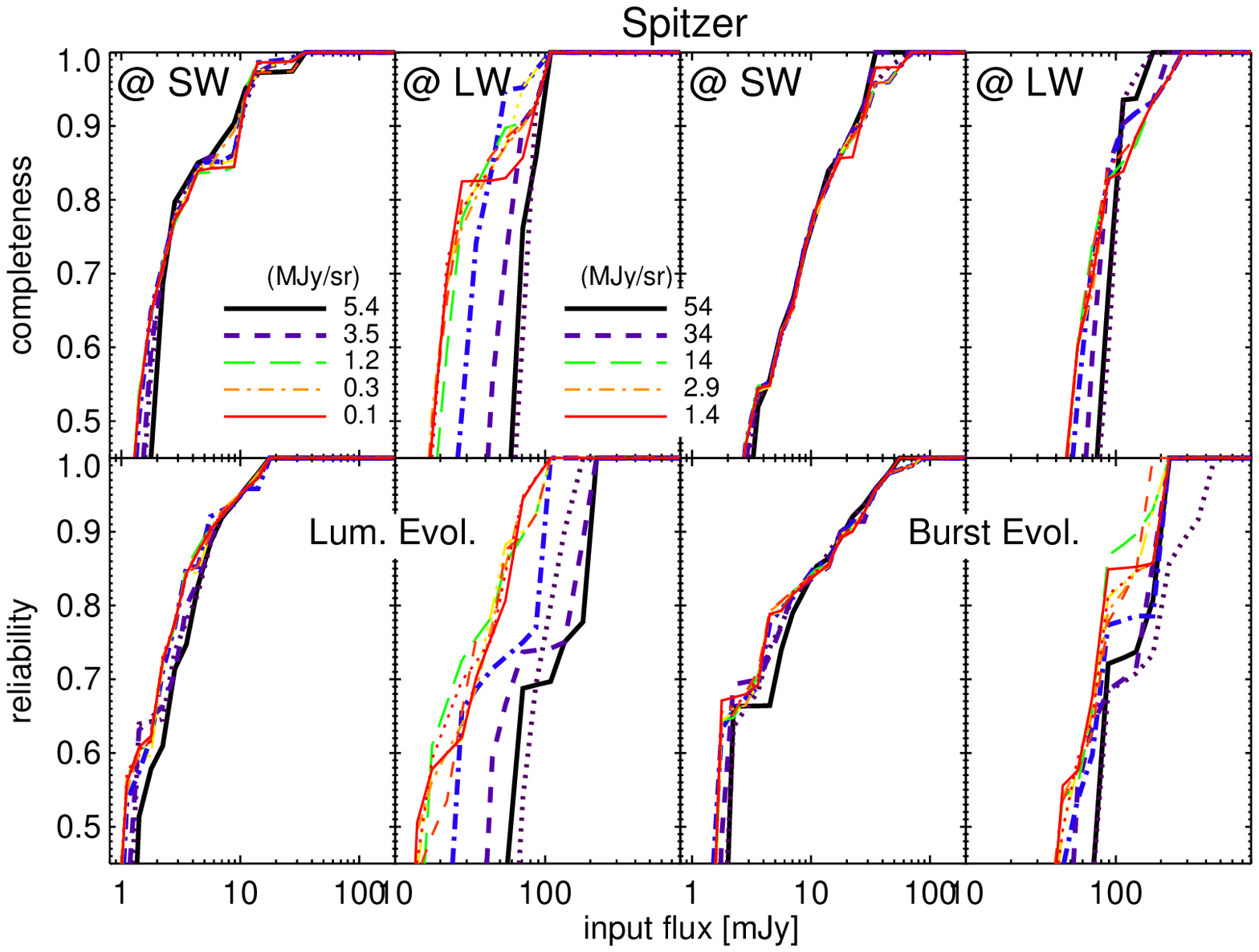}
    \end{center}
   \caption{Photometric results for the \textit{Spitzer} mission from the two evolutionary
   models. The explanations for each panel and lines are the same as in Fig.
   \ref{fig_phot_fis}. Compared to the case of the luminosity evolution model,
   the more severe source confusion makes each result similar in the case of
   burst evolution model especially in the LW band.}
   \label{fig_phot_sirtf_evols}
\end{figure*}

We also show the photometric results for the \textit{Spitzer} mission in the
case of the two evolutionary models in Fig. \ref{fig_phot_sirtf_evols}.
Although the cirrus background becomes higher, the completeness and reliability
show similar values in the case of both the SW and LW bands with a mean
brightness $\langle B_{\lambda}\rangle$ $<$ 10 MJy~sr$^{-1}$ due to the relatively
more severe source confusion compared to the case of the no evolution model. In
addition, we found that though the \textit{AKARI} and \textit{Spitzer}
missions have similar apertures (0.69m and 0.85m, respectively), the larger
pixel size of \textit{AKARI} makes both the sky and source confusion more
severe.

\begin{figure*}
  \begin{center}
    \epsfxsize = 13cm
    \epsffile{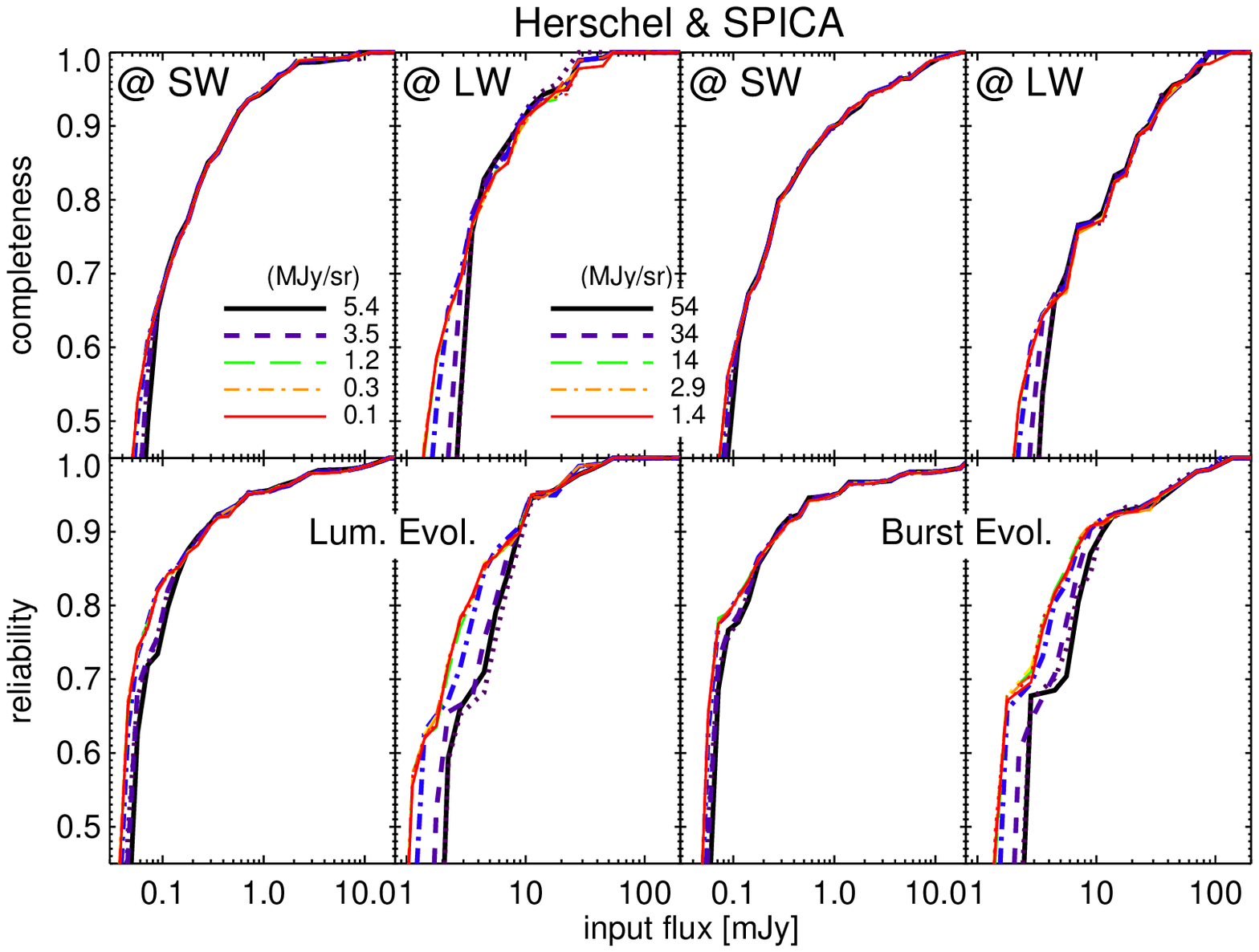}
    \end{center}
   \caption{Photometric results for the \textit{Herschel} \& \textit{SPICA} missions
   from the two evolutionary models. The explanations for each panel and lines are
   the same as in Fig. \ref{fig_phot_fis}. Since the source confusion is the dominant noise
   source, each result shows no significant difference, regardless of the
   cirrus background except for the case of the reliability in the LW band.}
   \label{fig_phot_spica}
\end{figure*}

Fig. \ref{fig_phot_spica} shows the photometric results for the
\textit{Herschel} and \textit{SPICA} missions from the two evolutionary models.
Since the source confusion is much more significant in most of the cirrus
background regions (see the lower panels of Fig. \ref{fig_conf_all}), the
completeness does not change significantly, irrespective of the amount of cirrus
background. However, from the reliability plot we see that the sky confusion still
affects the measurement of the source flux.

\section{ALL SKY CONFUSION MAPS} \label{app:conf_ass}

\begin{figure*}
  \begin{center}
    \epsfxsize = 8.5cm
    \epsffile{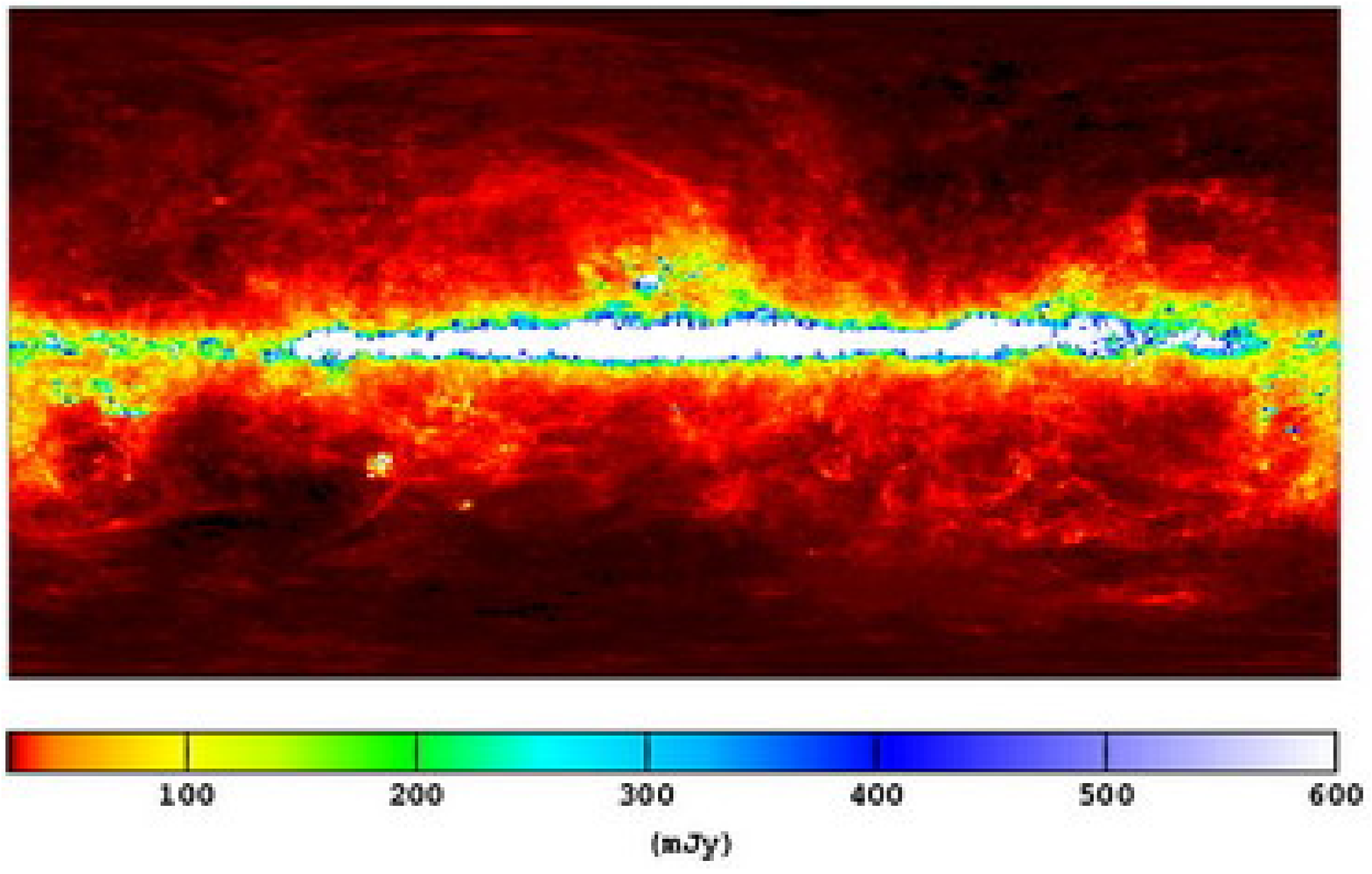}
    \hspace{10pt}
    \epsfxsize = 8.5cm
    \epsffile{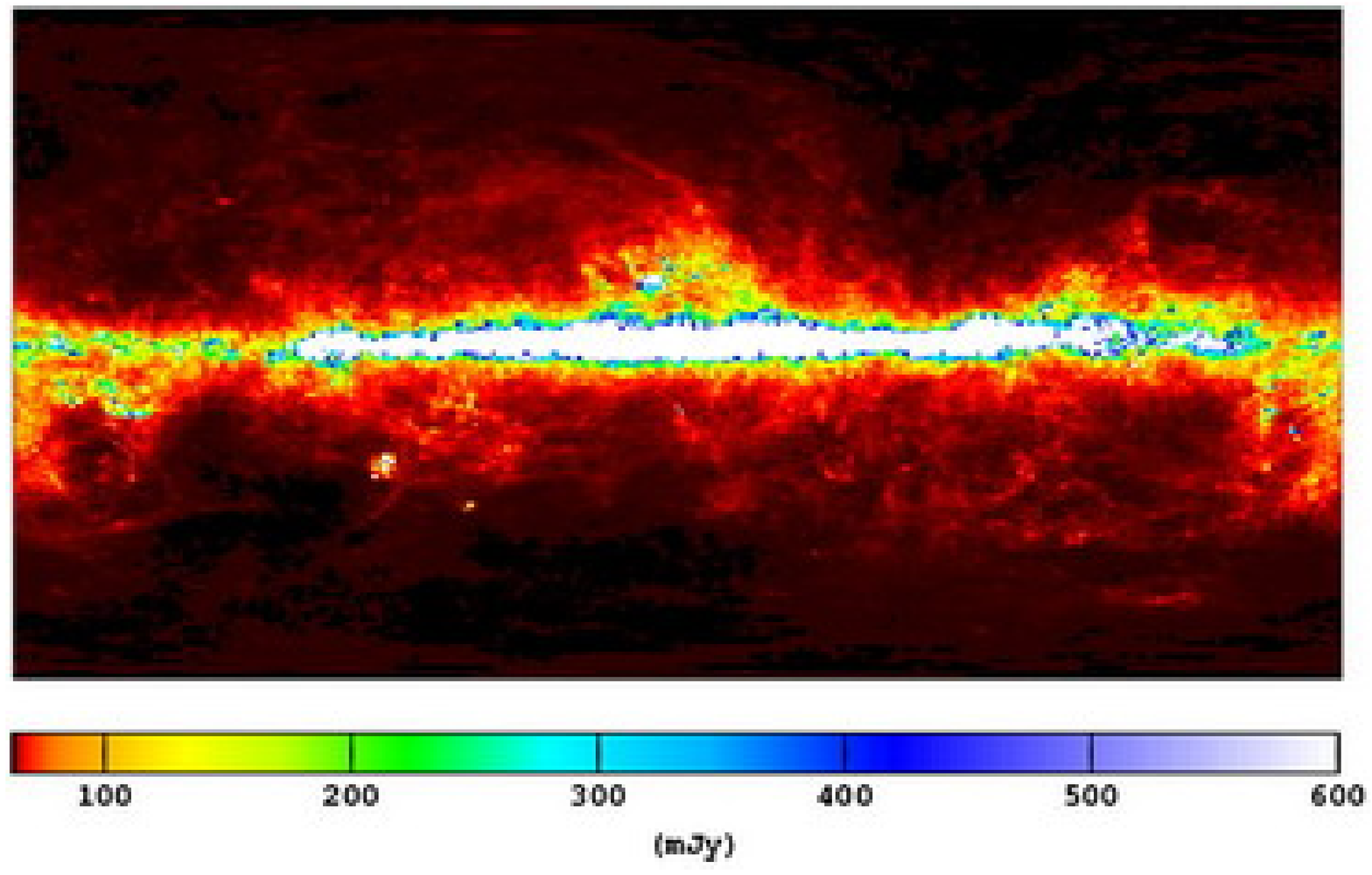}
    \end{center}
   \caption{Final confusion limit map for the \textit{Spitzer} mission
   in Galactic coordinates with the Galactic center located at the center
   of image. The left and right panels show the results for the luminosity
   evolution model and the burst evolution model, respectively.}
   \label{fig_conf_ass_sirtf}
\end{figure*}

\begin{figure*}
  \begin{center}
    \epsfxsize = 8.4cm
    \epsffile{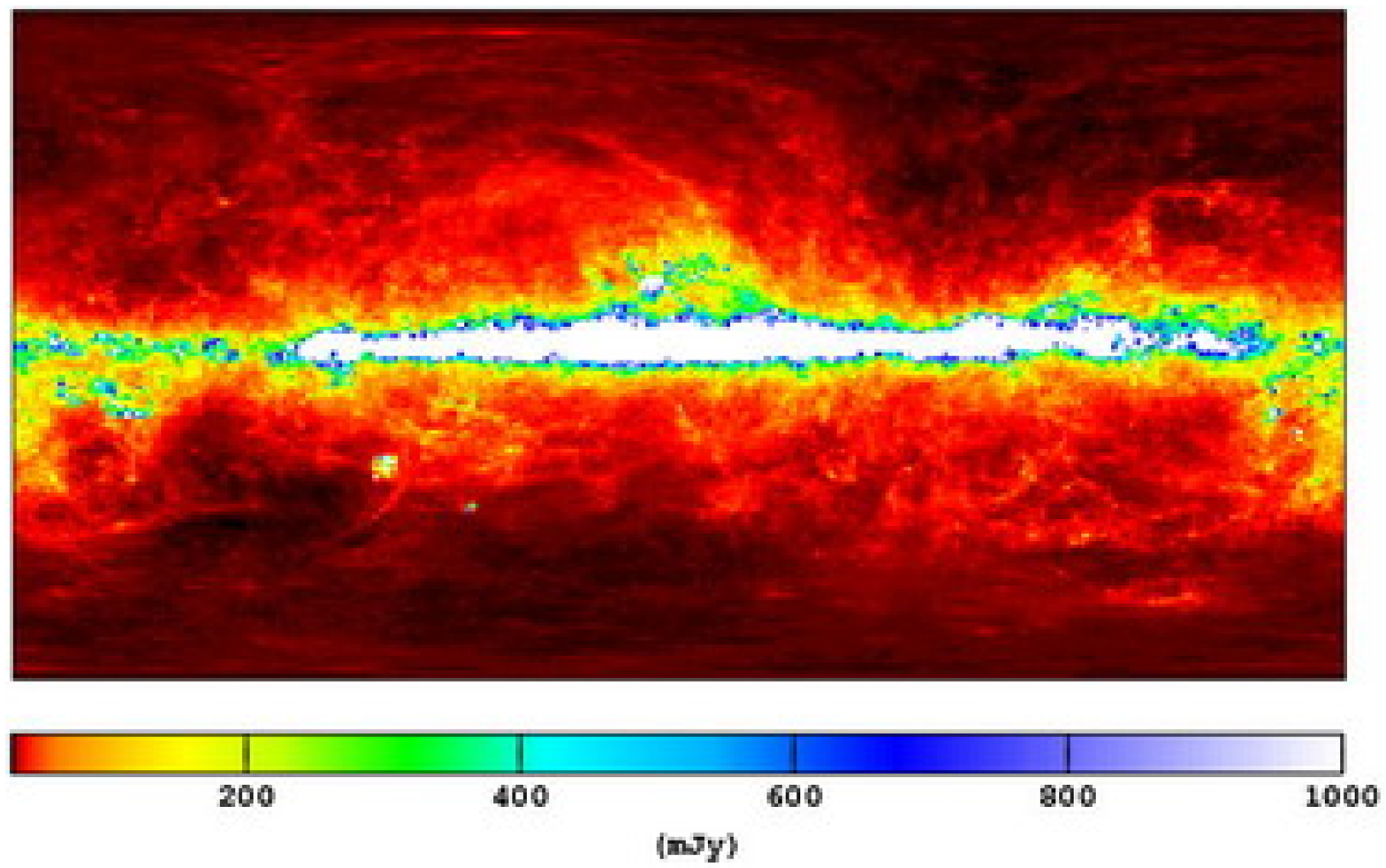}
    \hspace{10pt}
    \epsfxsize = 8.4cm
    \epsffile{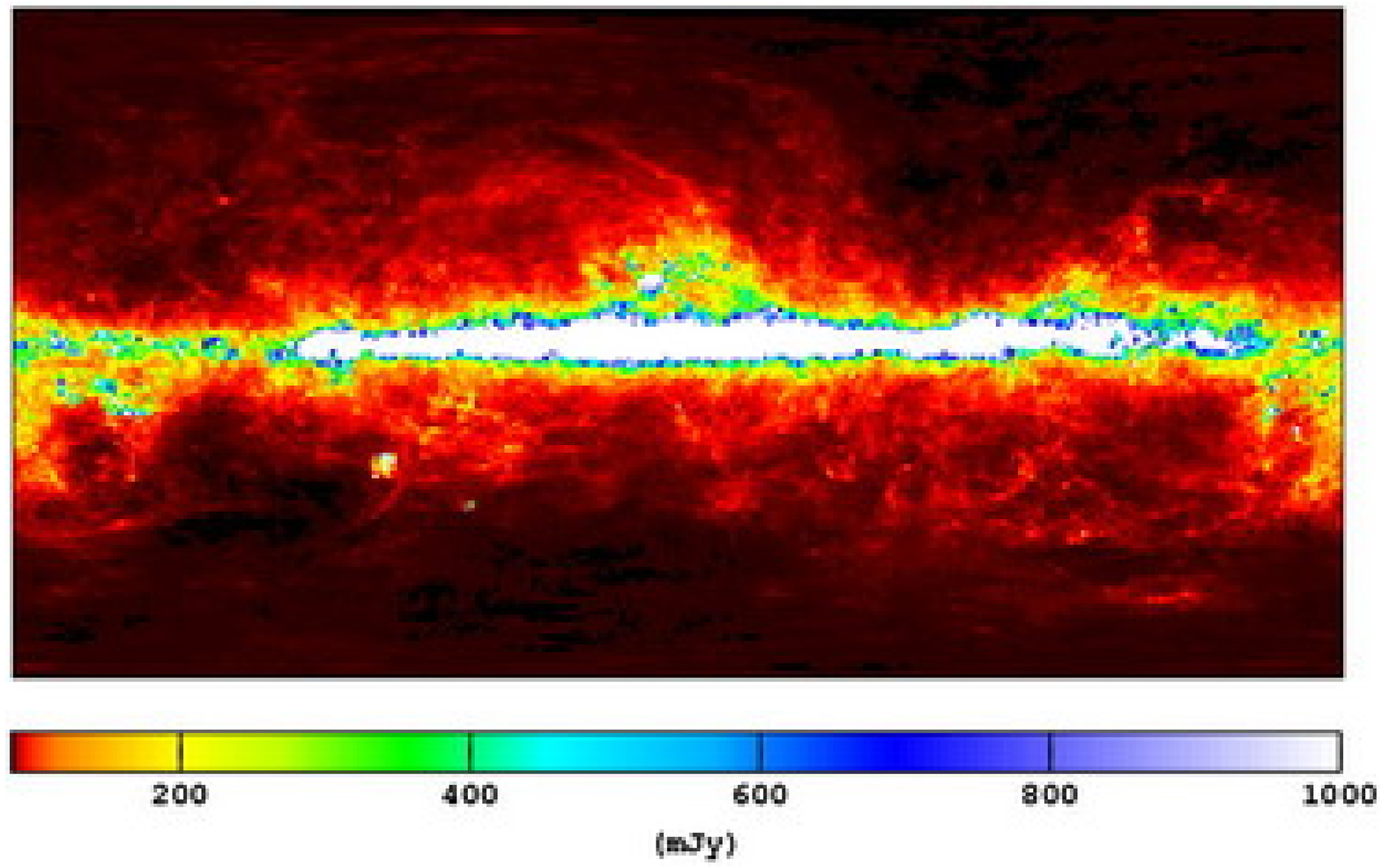}
    \end{center}
   \caption{Final confusion limit map for the scan mode observation
   of the \textit{AKARI} mission in Galactic coordinates with the Galactic center
   located at the center of image. The left and right panels show the results for the
   luminosity evolution model and the burst evolution model, respectively.}
   \label{fig_conf_ass_fis}
\end{figure*}

\begin{figure*}
  \begin{center}
    \epsfxsize = 8.4cm
    \epsffile{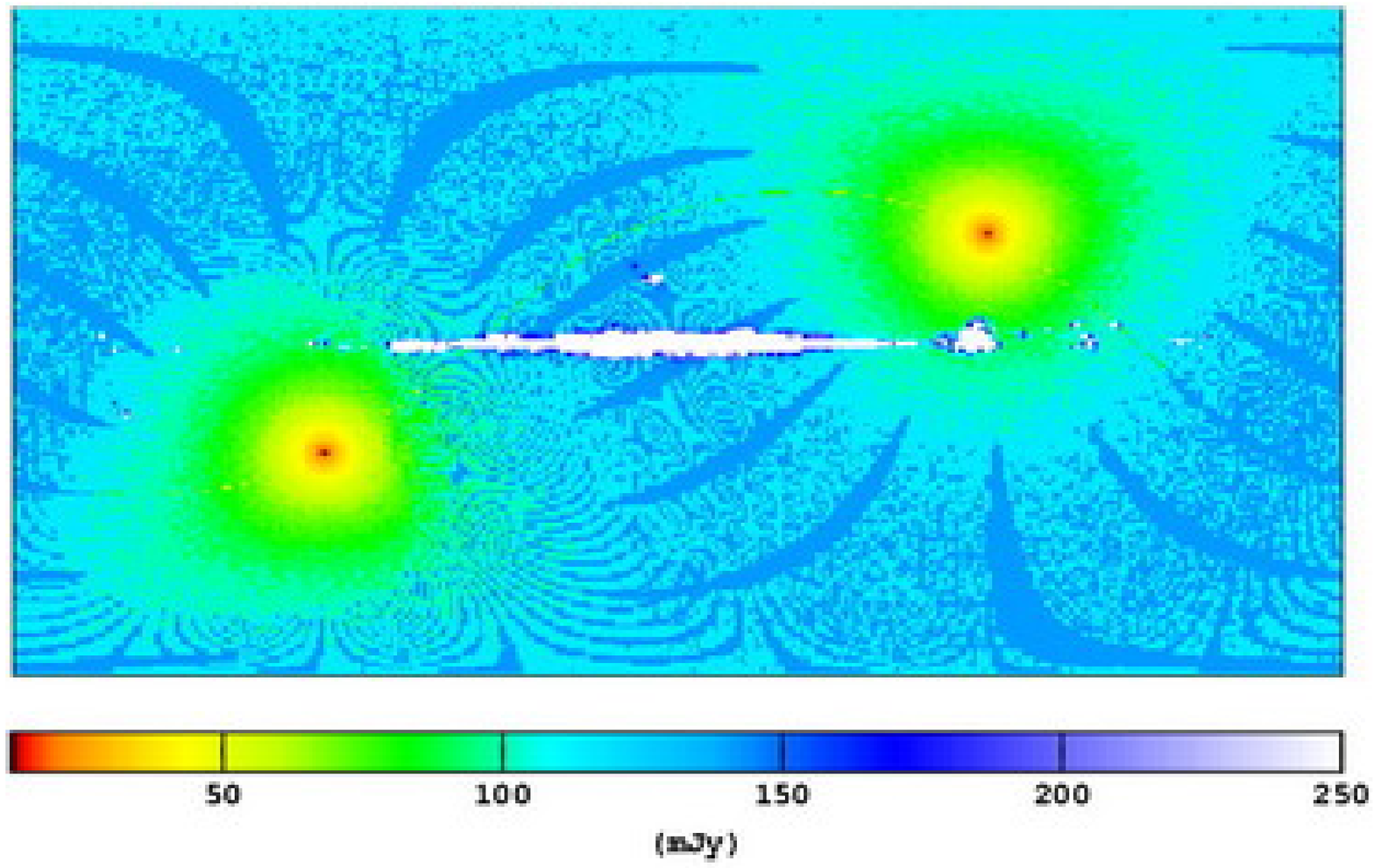}
    \hspace{10pt}
    \epsfxsize = 8.4cm
    \epsffile{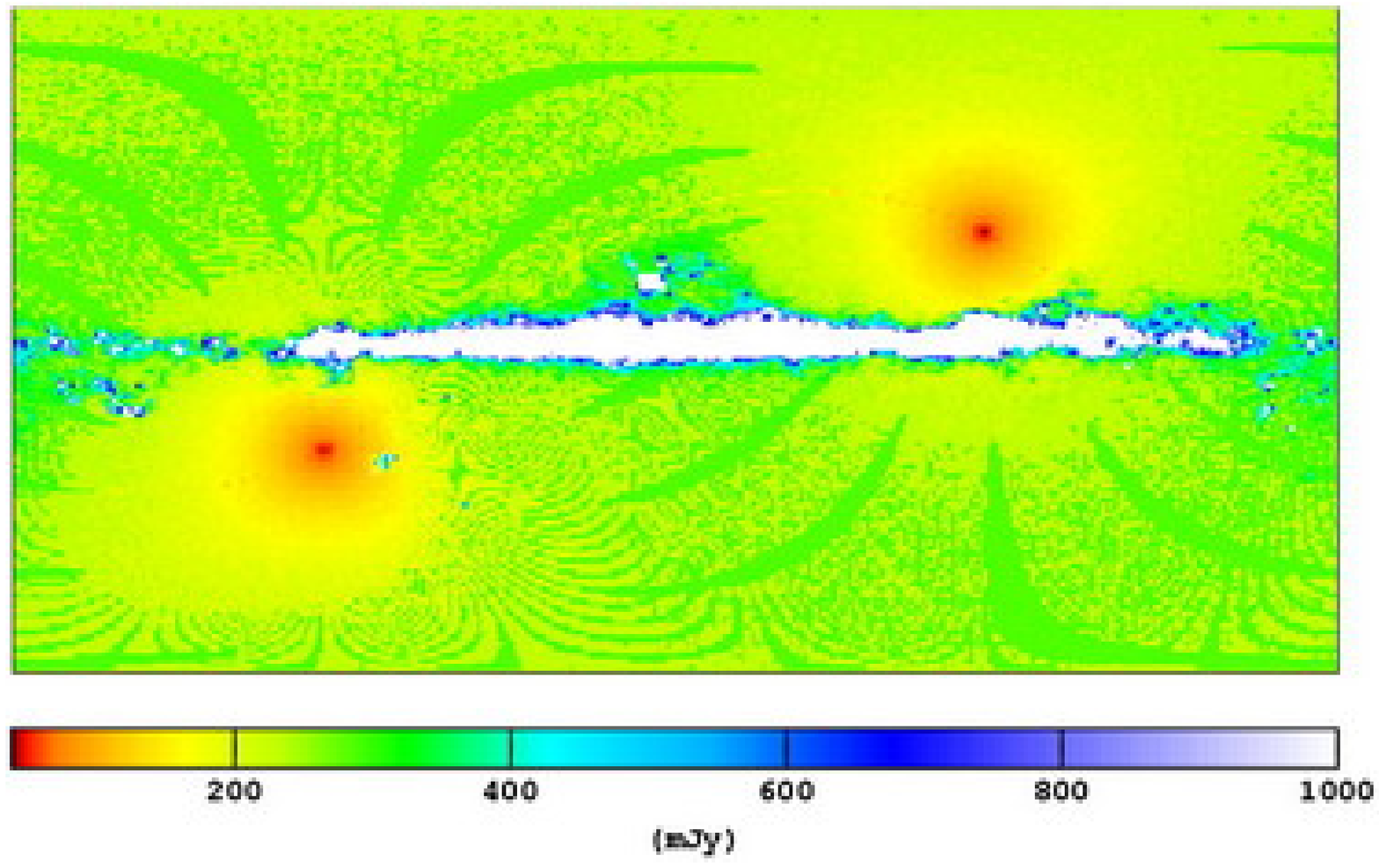}
    \end{center}
   \caption{Final detection limit map for the survey mode observation of
   the \textit{AKARI} mission in Galactic coordinates with the Galactic
   center located at the center of image. The left and right panels show the results
   for the SW band and the LW band, respectively. Since the detection limits mainly
   depend on the instrumental noise except for near the ecliptic poles and Galactic
   plane, we show only the case for the luminosity evolution model.}
   \label{fig_detlim_ass_fis}
\end{figure*}

\begin{figure*}
  \begin{center}
    \epsfxsize = 8.4cm
    \epsffile{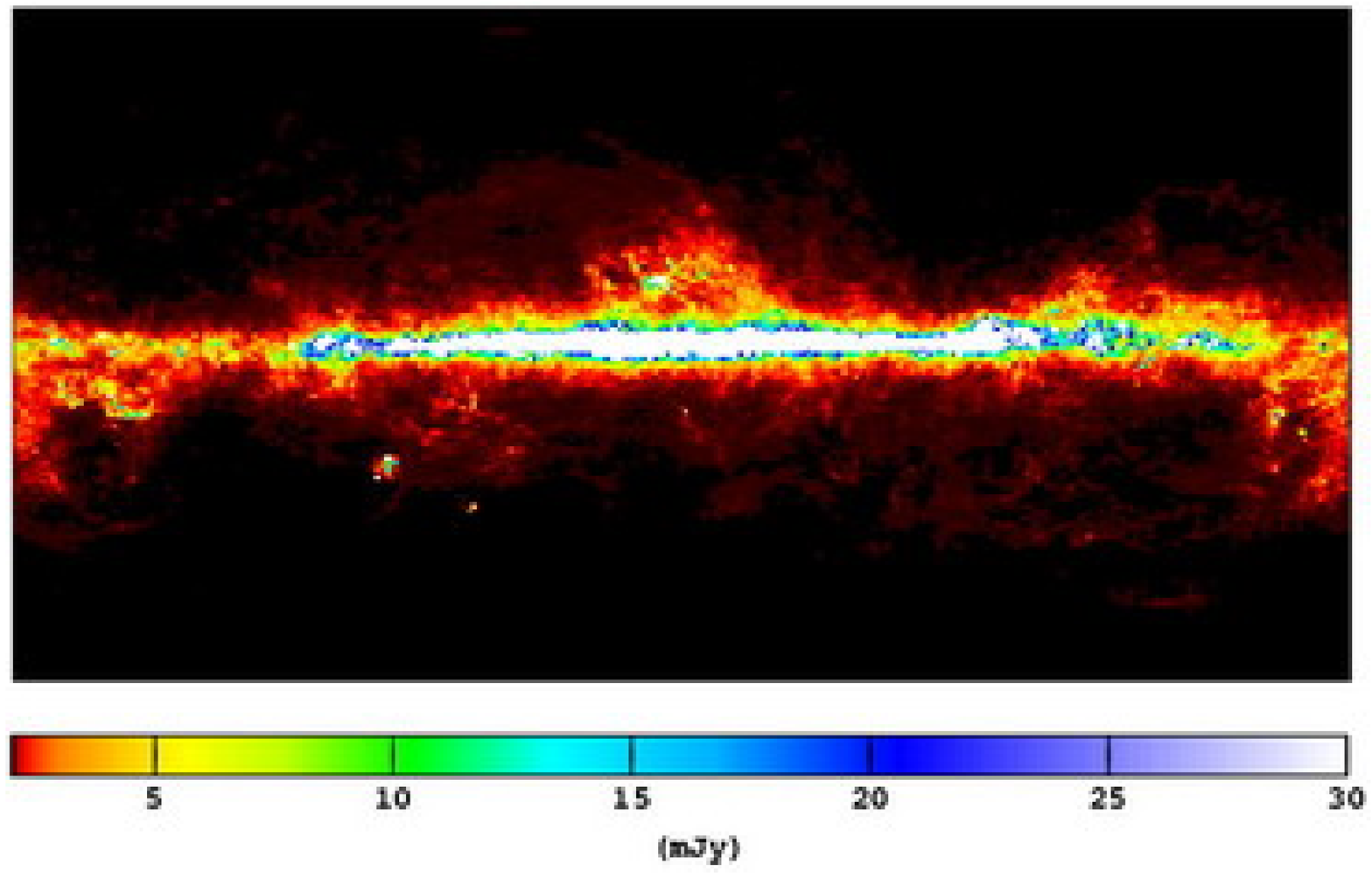}
    \hspace{10pt}
    \epsfxsize = 8.4cm
    \epsffile{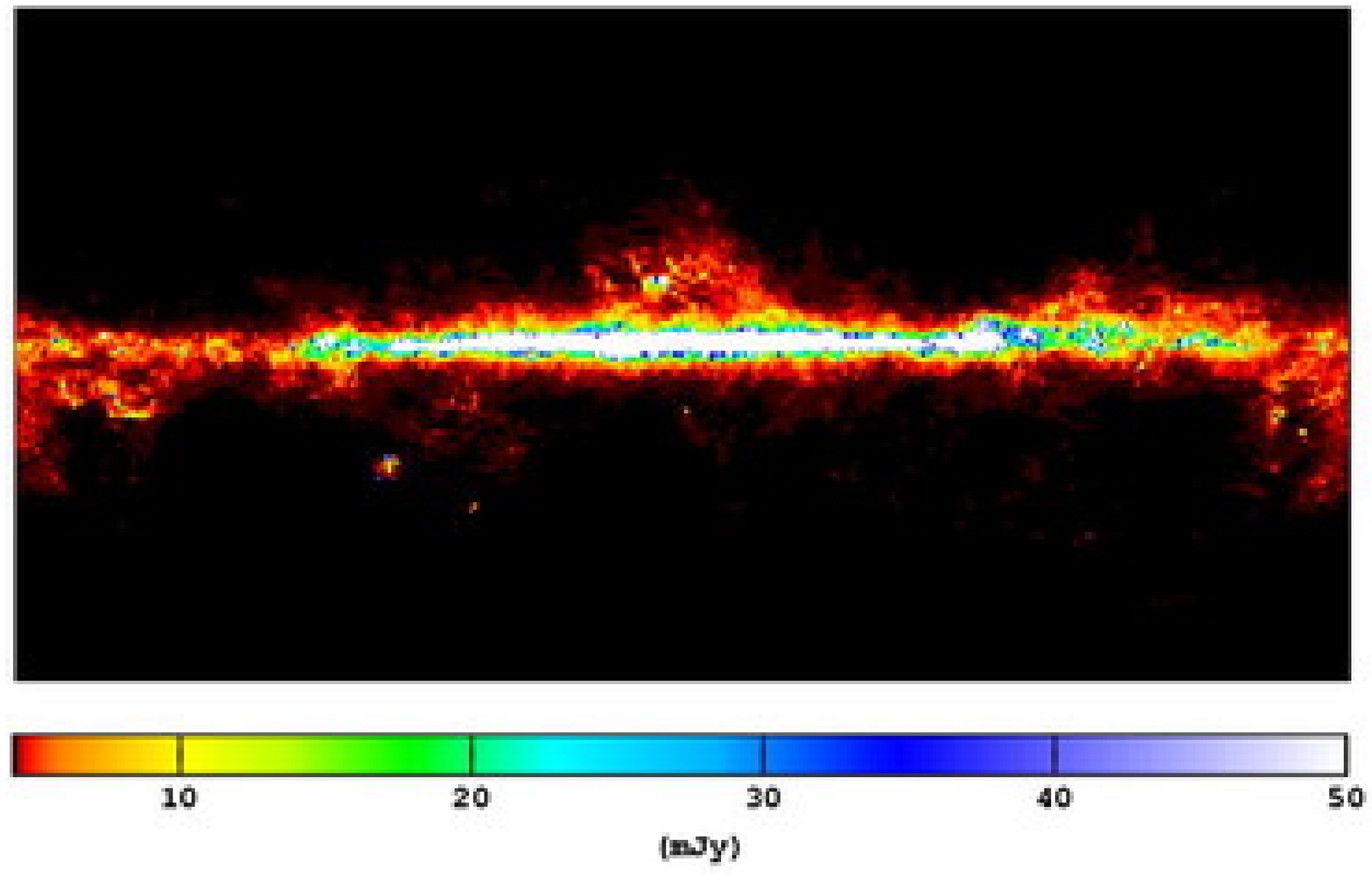}
    \end{center}
   \caption{Final confusion limit map for the \textit{Herschel}
   \& \textit{SPICA} missions in Galactic coordinates with the Galactic
   center located at the center of image. The left and right
   panels show the results  for the luminosity evolution model and the burst evolution model,
   respectively. Most of the regions are limited by source confusion.}
   \label{fig_conf_ass_spica}
\end{figure*}

Using our results for sky confusion and source confusion, we have generated all
sky maps for the final confusion limits. Since the final confusion limits in the SW
band are dominated by source confusion, we show our results only for the
case of the LW band (see Figs \ref{fig_conf_ass_sirtf}, \ref{fig_conf_ass_fis},
and \ref{fig_conf_ass_spica}) except for all sky survey of the \textit{AKARI}
mission. Since the instrumental noise of the all sky survey of the \textit{AKARI}
mission decreases in proportion to the square root of the number of scans, we
show here the final detection limit map including instrumental noise by
considering the visibility for the all sky survey\footnote{Further information
can be found at the following url:
\it{http://www.ir.isas.jaxa.jp/AKARI/Observation/vis/}} (see Figure
\ref{fig_detlim_ass_fis}).

\end{appendix}

\label{lastpage}

\end{document}